\documentclass{article}

\usepackage{PRIMEarxiv}

\usepackage[utf8]{inputenc} 
\usepackage[T1]{fontenc}    
\usepackage{hyperref}       
\usepackage{url}            
\usepackage{booktabs}       
\usepackage{amsfonts}       
\usepackage{nicefrac}       
\usepackage{microtype}      
\usepackage{lipsum}
\usepackage{fancyhdr}       
\usepackage{graphicx}       
\usepackage{float}
\usepackage{tabularx}
\usepackage{siunitx}
\usepackage{amsmath}
\usepackage{cite}
\graphicspath{{media/}}     

\title{Silicon and silicon nitride photonic ring resonator arrays for photonic trimming metrology  
}

\author{
  Cael Warner \\
  Electrical and Computer Engineering Department \\
  University of Alberta \\
  Edmonton\\
  spencerw@ualberta.ca}

\begin{document}
\maketitle

\begin{abstract}
Proposed silicon on silicon oxide 4$\times$4 matrix multiplier for use in refractive index measurements, based on the Advanced Micro Foundry deep ultraviolet process.  Agreement between simulated spectra and analytic design spectra demonstrate the intended performance of the components given a 1550 \si{nm} wavelength spectrum analysis. Intended purpose of the design includes optical damage analysis of the cladding, and photonic chip refractive index analysis following sub-threshold laser ablation. Femtosecond multi-shot tuning of refractive index is not a recommended application, due to round trip-losses causing quality factor and transfer function degradation.
\end{abstract}

\keywords{Matrix multiplier \and silicon photonics \and optical damage metrology \and silicon passives workshop}

\section{Introduction}
Silicon photonic devices promise rapid analog computing, particularly all-optical matrix operations in machine learning \cite{bogaerts2020programmable}, or localized measurements of refractive index change from incubation \cite{zhang2024precision}, non-thermal melting, strain \cite{schriever2012strained}, or ion implantation \cite{yu2022ge} which affect the crystalline structure and optical properties. In 2025, an all-passive silicon photonic matrix multiplier was designed from the Canadian Microsystems (CMC) silicon passives course (SiEPIC) following an Advanced Micro Foundry (AMF) process and process design kit (PDK). The first intended purpose of this design was for measuring cumulative resonance shift from real refractive index variation following repeated femto-second laser pulse ablation at 800 \si{nm} to 1030 \si{nm} wavelength \cite{zhanghigh}. However, this open-source design cannot satisfy current applications for patent of a femtosecond refractive index resonant shift \cite{UAlberta2025_003}, such that a secondary AMF photonic circuit was adopted for use in the AMF process \cite{zhanghigh}. Programmable photonic circuits typically have strict quality factor and signal requirements \cite{bogaerts2020programmable}, which may not be satisfied from nonuniform non-thermal heating \cite{warner2025simulation} and incubation which contribute to Rayleigh and Mie scattering within the silicon lattice of the waveguide \cite{bachman2016postfabrication}.

In this work, the layout and expected design performance of the photonic circuit components are described in detail, including ANSYS\textsuperscript{\textregistered} Lumerical 2025 three-dimensional (3D) finite-difference time-domain (FDTD) and varFDTD simulations. Simulations demonstrate a broad-band variation in the signal transmission, such that round-trip losses are critical to the transfer function of the multi-microring photonic arrays, which are not recommended for multi-shot femtosecond infrared pulse tuning directly to the silicon \cite{zhang2024precision}. Alternatively, we strongly recommend infrared multi-shot tuning of the silicon oxide cladding (such as by Femtum Inc.), ion or proton implantation \cite{yu2022ge}, or Dream Photonics Fabrication-Aware Inverse Design (FAID), which do not require local micro cracks or lattice distortions degrading signal quality to shift the resonance frequency.

\newpage
\section{Theory}

In this section, analytic framework for microring resonators, three-ring Butterworth filters, and matrix multipliers are reviewed prior to a discussion into anticipated uses of this photonic circuit for monitoring optical damage or refractive index tuning techniques. Finally, a disclaimer is included, where alternative techniques from femtosecond ablation are recommended for this particular photonic circuit layout and process.

\subsection{Silicon photonic ring resonators}
Silicon photonic ring resonators transmit light confined within a rectangular waveguide, also referred to as a strip waveguide, which is arced into a particular radius circle with an optical path length that is an integer multiple of the vacuum wavelength

\begin{equation}
    m\lambda_m = n_\mathrm{eff} L = n_\mathrm{eff}(2\pi r),
\end{equation}
where $m=\{1,2,3, ...|  m\in \mathbb{Z}\}$, $n_\mathrm{eff}$ is the effective refractive index of the ring waveguide mode, $L$ is the round-trip circumference $L=2\pi r$, and $r$ is the ring radius. Free-spectral range (FSR) of the microring is
\begin{equation}
    \mathrm{FSR}_\lambda = \frac{\lambda_m^2}{n_g L},\quad\mathrm{or}\quad \mathrm{FSR}_\nu = \frac{c}{n_gL},
\end{equation}
where $n_g = n_\mathrm{eff} - \lambda \mathrm{d} n_\mathrm{eff}/\mathrm{d} \lambda$, and $c$ is the speed of light. In a single bus waveguide coupled to a ring with self-coupling coefficient $a_1$ (amplitude transmission across coupling) and round trip attenuation factor $a=\exp(-\alpha L/2)$
\begin{equation}
    T(\theta) = \left|\frac{a_1-ae^{i\theta}}{1-a_1ae^{i\theta}}\right|^2 = \frac{a_1^2 + a^2 -2a_1a\cos(\theta)}{1+a_1^2a^2-2a_1a\cos(\theta)},
\end{equation}
where $\theta = 2\pi n_\mathrm{eff} L/\lambda$ is the single-pass phase shift. Interference between the receiving wave and transmitting wave results in resonance, yet relies on a particular coupling where critical coupling $a_1=a$ would result in zero transmission; under-coupling $a_1>a$ has weaker coupling than internal loss; and over-coupling has greater coupling than internal loss. The corresponding quality factor measures the energy storage in the ring relative to the power loss per cycle
\begin{equation}
Q = \frac{\lambda_0}{\Delta \lambda_\mathrm{FWHM}} = \frac{\nu_0}{\Delta \nu_\mathrm{FWHM}},    
\end{equation}
where $\lambda_0$ is the resonant wavelength and $\Delta \lambda_\mathrm{FWHM}$ is the full-width at half-maximum spectral linewidth of the resonance spectrum. With respect to intrinsic and coupling losses, the loaded quality factor is
\begin{equation}
    \frac{1}{Q_\mathrm{loaded}} = \frac{1}{Q_\mathrm{int}} + \frac{1}{Q_\mathrm{ext}}.
\end{equation}
In the case of an all-pass microring
\begin{equation}
    Q_\mathrm{int} = \frac{2\pi n_g}{\lambda_0 \alpha} = \frac{\pi n_g L}{\lambda_0(1-a^2)},\quad \mathrm{and}\quad Q_\mathrm{ext} = \frac{\pi n_g L}{\lambda_0(1-a_1^2)}.
\end{equation}
Finally, finesse relates the FSR to resonance width
\begin{equation}
    \mathcal{F} = \frac{\mathrm{FSR}}{\Delta \lambda_\mathrm{FWHM}} = \frac{\pi \sqrt{a_1 a}}{1-a_1a},
\end{equation}
and is related to the quality factor via
\begin{equation}
    Q = m\mathcal{F}.
\end{equation}
For an add-drop configuration with field self-coupling coefficients $r_1, r_2$ at the input and drop couplers respectively, single-pass power attenuation factor $a = e^{-\alpha L / 2}$, and single-pass phase shift $\theta = \frac{2\pi n_{\text{eff}} L}{\lambda}$. The corresponding transmission of the through port is

\begin{equation}
T_{\text{through}}(\theta) = \left| \frac{r_1 - r_2 a e^{i\theta}}{1 - r_1 r_2 a e^{i\theta}} \right|^2 = \frac{r_1^2 + r_2^2 a^2 - 2 r_1 r_2 a \cos\theta}{1 + r_1^2 r_2^2 a^2 - 2 r_1 r_2 a \cos\theta},
\end{equation}
whereas the transmission through the drop port is

\begin{equation}
T_{\text{drop}}(\theta) = \left| \frac{-t_1 t_2 \sqrt{a} e^{i\theta/2}}{1 - r_1 r_2 a e^{i\theta}} \right|^2 = \frac{(1 - r_1^2)(1 - r_2^2) a}{1 + r_1^2 r_2^2 a^2 - 2 r_1 r_2 a \cos\theta}
\end{equation}

where $t_1, t_2$ represent cross-coupling amplitudes ($r_i^2 + t_i^2 = 1$). Critical coupling conditions occur at resonance ($\theta = 2m\pi$), when the through-port power transmission drops to zero ($T_{\text{through}} = 0$) such that

\begin{equation}
r_1 = r_2 a.
\end{equation}

In symmetric coupling ($r_1 = r_2 = r$), zero transmission at the through port occurs only in an ideal lossless ring ($a = 1$). In the case of a lossy cavity ($a < 1$), symmetric coupling yields an under-coupled response. Finally, the loaded quality factor $Q_{\text{loaded}}$ accounts for internal absorption/scattering and external escape rates into both buses:

\begin{equation}
\frac{1}{Q_{\text{loaded}}} = \frac{1}{Q_{\text{int}}} + \frac{1}{Q_{\text{ext1}}} + \frac{1}{Q_{\text{ext2}}},
\end{equation}

with individual quality factors for each port

\begin{align}
Q_{\text{int}} &= \frac{2\pi n_{\text{g}}}{\lambda_0 \alpha} \approx \frac{\pi n_{\text{g}} L}{\lambda_0 (1 - a^2)}, \\
Q_{\text{ext1}} &= \frac{\pi n_{\text{g}} L}{\lambda_0 (1 - r_1^2)} \\
Q_{\text{ext2}} &= \frac{\pi n_{\text{g}} L}{\lambda_0 (1 - r_2^2)}.
\end{align}

In the case of a symmetric filter ($r_1 = r_2 = r$) with negligible internal loss ($a \approx 1$)

\begin{equation}
Q_{\text{loaded}} \approx \frac{\pi n_{\text{g}} L}{2 \lambda_0 (1 - r^2)}.
\end{equation}

\begin{figure}[H]
    \centering    \includegraphics[width=0.8\linewidth]{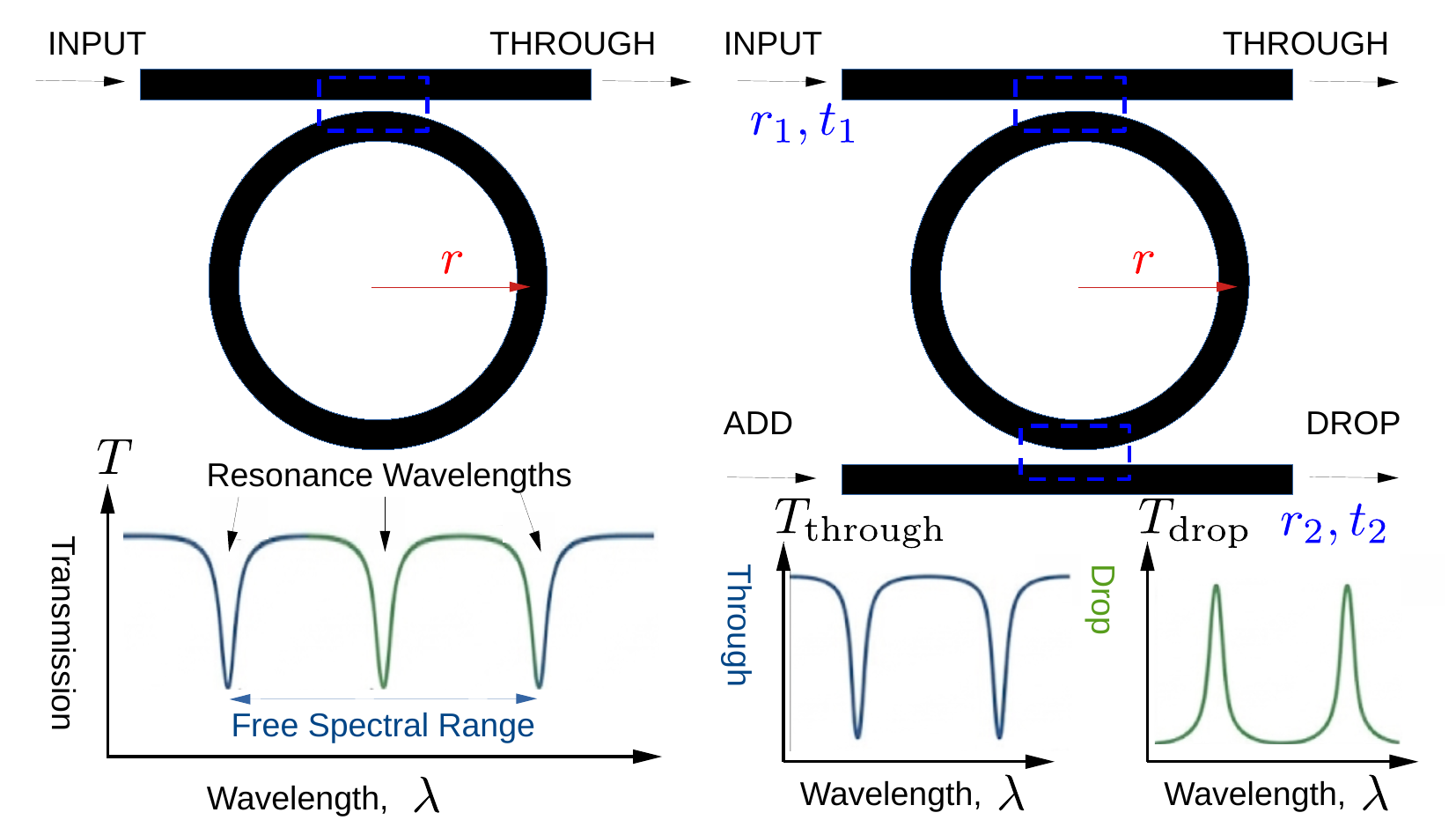}
    \caption{All-pass and add-drop configurations.}
    \label{fig:placeholder}
\end{figure}

\newpage
\subsection{Third-order coupled Butterworth filter resonator/modulator} 

\begin{figure}[H]
    \centering    \includegraphics[width=0.5\linewidth]{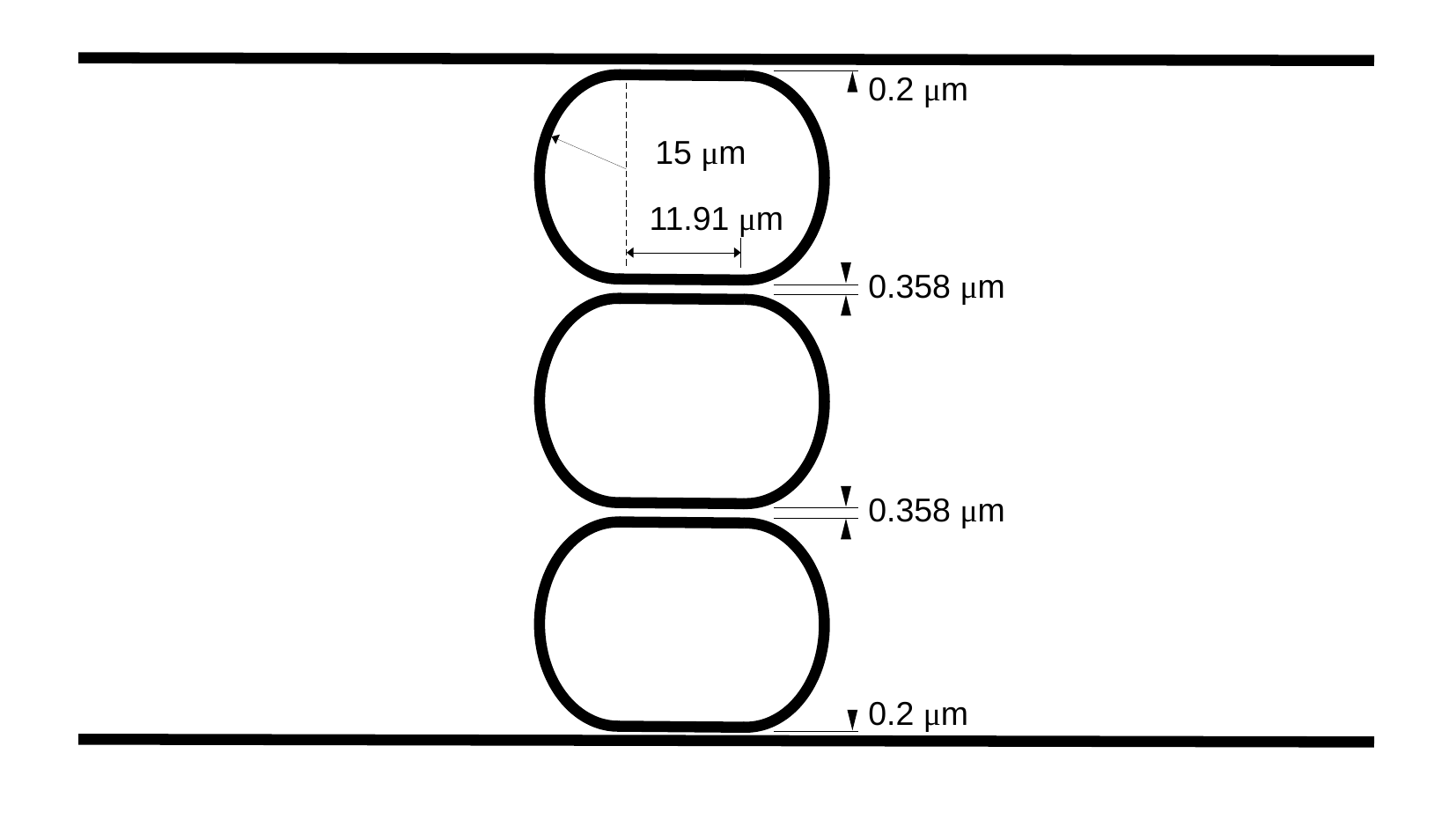}
    \caption{Three-stage ring resonator analytical model diagram. Design ideation from Robin Kim.}
    \label{fig:5}
\end{figure}

Figure \ref{fig:5} shows an annotated diagram of the third-order resonator/modulator analytical model and the expected spectrum given ideal fabrication.

\begin{figure}[H]
    \centering    \includegraphics[width=0.42\linewidth]{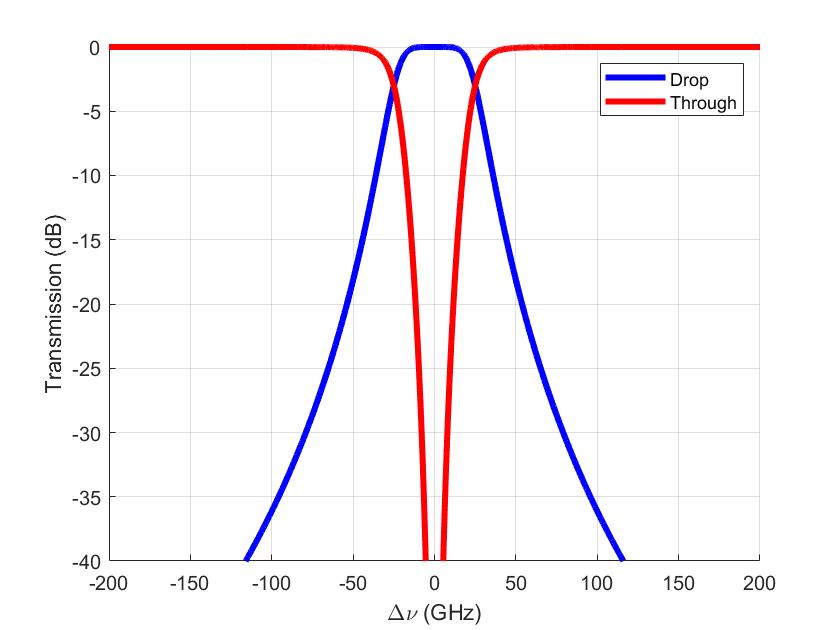}
       \includegraphics[width=0.42\linewidth]{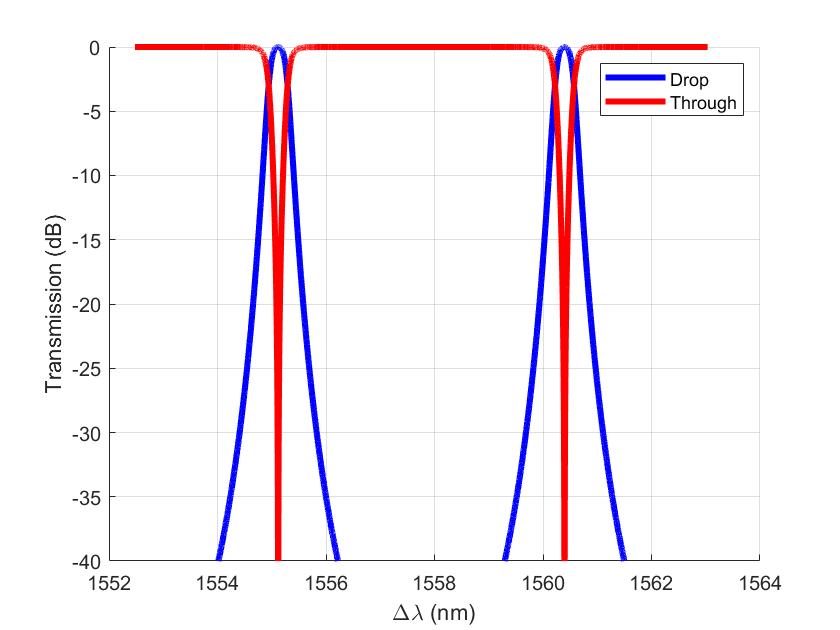}
    \caption{Through and drop port transfer function for $\Delta \nu_B =50$ \si{GHz} Butterworth filter. Design ideation from Robin Kim.}
    \label{fig:6}
\end{figure}

The transfer function of the third-order serially coupled racetrack resonator array (CROW) is a flat-top Butterworth filter. This transfer function may be derived from the energy coupling formalism to realize an analog filter in the s-domain. In the ideal analytical solution, it is assumed that each racetrack is identical and lossless ($\gamma_0 = 0$). Therefore, the drop port transfer function obtained from the coupled mode equations has the form
\begin{equation}
    H_d(s) = \frac{s_d}{s_i} = \frac{\mu_0\mu_1\mu_2\mu_3}{C_3(s)}
\end{equation}
where $s = j\Delta\omega$ ($\Delta \omega$ is the frequency detuned from resonance), $\mu_i$ is the energy coupling coefficient between the waveguides ($\mu_0$ denotes coupling from input bus waveguide to the first racetrack, $\mu_1$ denotes coupling from the first and second racetrack, and so on). The function $C_3(s)$ is a third-order polynomial
\begin{equation}
    C_3(s) = \det(\mathbf{M}) = 
\begin{vmatrix}
s+\mu_0^2/2 & j\mu_1 & 0 \\
j\mu_1 & s & j\mu_2 \\
0 & j\mu_2 & s+\mu_3^2/2
\end{vmatrix},
\end{equation}
\begin{equation}
    C_3(s)=s^3+(\frac{\mu_0^2+\mu_3^2}{2})s^2+(\mu_1^2+\mu_2^2+\frac{\mu_0^2\mu_3^2}{4})s+\frac{\mu_2^2\mu_0^2+\mu_1^2\mu_3^2}{2}.
\end{equation}
Through and drop port transmission are plotted in Figure \ref{fig:6} for the case of a third-order Butterworth filter with a \mbox{-}3\si{dB} bandwidth of $\Delta \nu_{B} = 50$ \si{GHz} and energy coupling coefficients $\mu_0 = \mu_3 = 17.7$ \si{GHz^{0.5}} and $\mu_1 = \mu_2 = 111.1$ \si{GHz}. The transfer function from the power coupling model is assumed to be a periodic transfer function of the energy coupling model. Gaps and racetrack lengths correspond to the targeted coupling coefficients obtained using supermode analysis (simulations performed using the beam propagation method or BPM). The design parameters are listed in Table \ref{tab:1}.

\subsection{Silicon ring through-port matrix multiplier}

\begin{figure}[H]
    \centering    \includegraphics[width=0.8\linewidth]{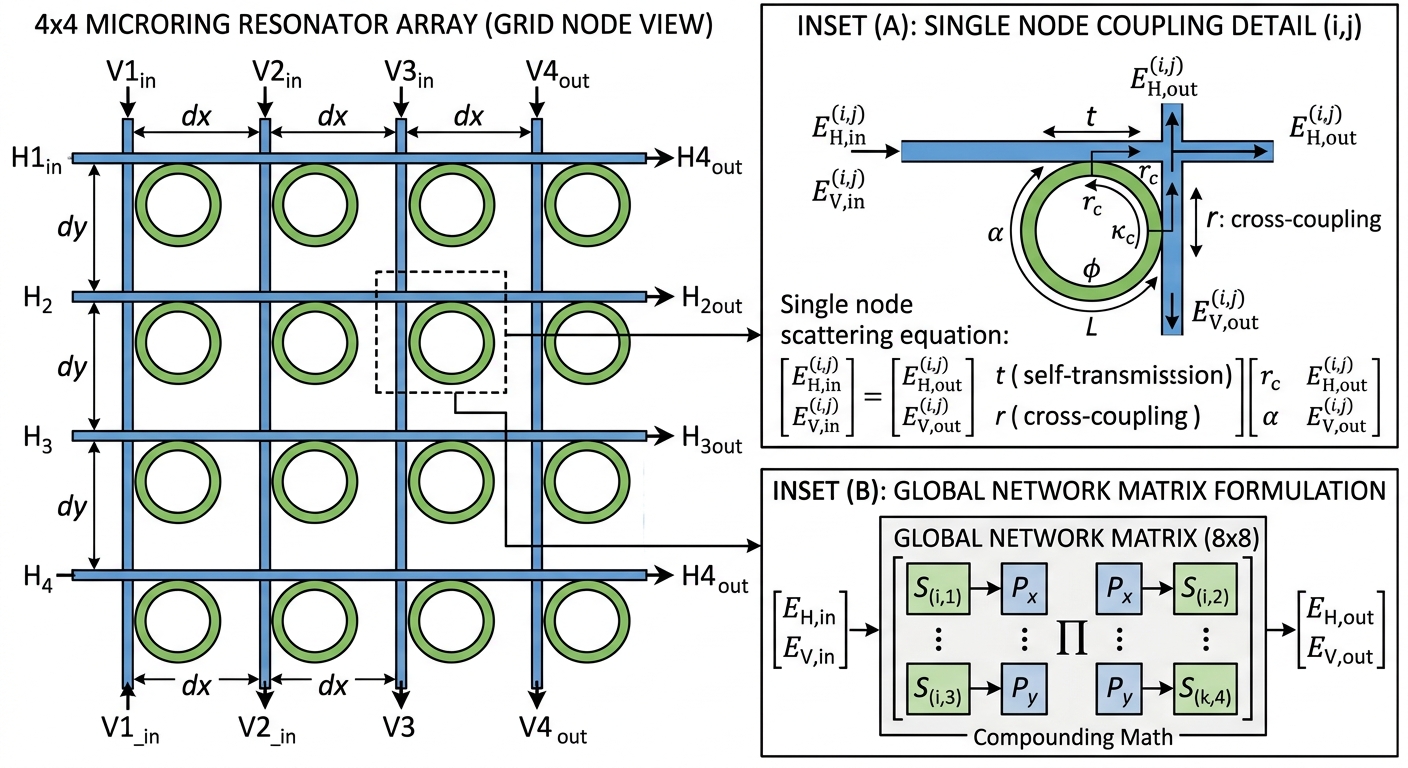}
    \caption{Ring resonator based 4$\times$4 matrix multiplier with orthogonal waveguides as a 4-port junctions. Image was generated using Gemini Flash 3.6.}
    \label{fig:ring_resonator_diagram}
\end{figure}

Matrix multiplier optical element configurations are well-established in standard optical photonics literature \cite{bogaerts2020programmable,madsen1999optical}.
Consider a grid formed by 4 horizontal bus waveguides ($H_1, \dots, H_4$) and 4 vertical bus waveguides ($V_1, \dots, V_4$). At each intersection $(i, j)$ for $i, j \in \{1, 2, 3, 4\}$, a microring resonator couples the orthogonal waveguides as a 4-port junction. Fig. \ref{fig:ring_resonator_diagram} for an illustration.

The scattering relation for a single node $(i,j)$ relates incoming field amplitudes to outgoing field amplitudes as

\begin{equation}
\begin{pmatrix} E_{H,\text{out}}^{(i,j)} \\ E_{V,\text{out}}^{(i,j)} \end{pmatrix} = \mathbf{S}_{\text{node}} \begin{pmatrix} E_{H,\text{in}}^{(i,j)} \\ E_{V,\text{in}}^{(i,j)} \end{pmatrix}.
\end{equation}

In the case of a symmetric microring coupler, the scattering matrix is

\begin{equation}
\mathbf{S}_{\text{node}}(\phi) = \begin{pmatrix} t(\phi) & r(\phi) \\ r(\phi) & t(\phi) \end{pmatrix}.
\end{equation}

where the self-transmission coefficient $t(\phi)$ and cross-coupling coefficient $r(\phi)$ around resonance are

\begin{equation}\label{eq:trans_cpling}
t(\phi) = \frac{r_c - a e^{i\phi}}{1 - r_c^2 a e^{i\phi}}, \quad\textrm{and}\quad r(\phi) = \frac{-\kappa_c^2 \sqrt{a} e^{i\phi/2}}{1 - r_c^2 a e^{i\phi}}.
\end{equation}

In Eq. (\ref{eq:trans_cpling}):
\begin{itemize}
    \item $r_c, \kappa_c$ are the self- and cross-coupling field coefficients of the directional couplers ($r_c^2 + \kappa_c^2 = 1$);
    \item $a = e^{-\alpha L / 2}$ is the single-pass field amplitude attenuation ($L = 2\pi r$); and
    \item $\phi = 2\pi n_{\text{eff}} L/\lambda$ is the single-pass optical phase shift of the ring.
\end{itemize}

Between adjacent intersections along the grid, spatial phase of the light accumulates along horizontal ($x$) and vertical ($y$) bus channels as

\begin{equation}
\mathbf{P}_x = \begin{pmatrix} e^{-i \theta_x} & 0 \\ 0 & e^{-i \theta_x} \end{pmatrix}, \quad \mathbf{P}_y = \begin{pmatrix} e^{-i \theta_y} & 0 \\ 0 & e^{-i \theta_y} \end{pmatrix}
\end{equation}

where $\theta_x = \beta \cdot d_x$, $\theta_y = \beta \cdot d_y$, $\beta = 2\pi n_{\text{eff, bus}}/\lambda$ is the propagation constant of the bus waveguide, and $d_x, d_y$ are inter-ring node distances.

Combining all inputs into global vectors $\mathbf{E}_H = [E_{H1}, E_{H2}, E_{H3}, E_{H4}]^\mathrm{T}$ and $\mathbf{E}_V = [E_{V1}, E_{V2}, E_{V3}, E_{V4}]^\mathrm{T}$, the full 8-port output response is governed by:

\begin{equation}
\begin{pmatrix} \mathbf{E}_H^{\text{out}} \\ \mathbf{E}_V^{\text{out}} \end{pmatrix} = \mathbf{M}_{\text{array}}(\lambda) \begin{pmatrix} \mathbf{E}_H^{\text{in}} \\ \mathbf{E}_V^{\text{in}} \end{pmatrix}.
\end{equation}

The corresponding array scattering matrix $\mathbf{M}_{\text{array}}$ is evaluated by cascading node scattering matrices $\mathbf{S}_{(i,j)}$ and propagation matrices $\mathbf{P}_x, \mathbf{P}_y$ using Redheffer star products

\begin{equation}
\mathbf{M}_{\text{array}} = \prod_{i=1}^4 \prod_{j=1}^4 \mathbf{T}_{(i,j)}(\phi_{i,j}, \theta_x, \theta_y).
\end{equation}

When signals enter solely through horizontal ports ($\mathbf{E}_V^{\text{in}} = \mathbf{0}$), the system simplifies to a linear matrix transformation mapping inputs to throughput and drop outputs as

\begin{equation}
\mathbf{E}_H^{\text{out}} = \mathbf{T}_{\text{through}} \mathbf{E}_H^{\text{in}}, \quad \textrm{and}\quad \mathbf{E}_V^{\text{out}} = \mathbf{T}_{\text{drop}} \mathbf{E}_H^{\text{in}}.
\end{equation}

Each component $T_{i,j}(\lambda)$ of the transfer matrix equals the sum over all coherent multi-path optical routes connecting input port $i$ to output port $j$

\begin{equation}
T_{i,j}(\lambda) = \sum_{p \in \text{paths}} A_p \exp\left(i \sum \phi_{\text{path}}\right).
\end{equation}

Individual phase shifters $\phi_{i,j}$ can be controlled at each node, such that the power transfer matrix elements $|T_{i,j}|^2$ can be dynamically programmed to execute optical matrix-vector multiplications ($\mathbf{y} = \mathbf{A}\mathbf{x}$).

\subsection{Limitation of purpose for non-thermal melting}

Matrix multipliers based on silicon ring resonators through ports and drop ports can be simultaneously measured during resonance. Simultaneous measurements of through-port and drop-port resonances allow local measurement of resonance shifts and refractive index changes within the rings. Therefore, such a passive silicon device has a secondary application for monitoring localized optical damage.

Low-repetition rate 800 \si{nm} wavelength femtosecond ablation \cite{zhang2023study,zhang2024multi,zhang2024precision,zhanghigh}, which results in two-photon absorption and an incubation effect \cite{zhang2023study}, has been proposed as an irreversible approach to tune refractive index resonant frequency as a post-fabrication photonic trimming technique. Whereas this approach shifts the resonance frequency \cite{zhanghigh}, it degrades a cavity's quality factor ($Q$). Multi-shot sub-threshold incubation drives non-equilibrium phase transitions in crystalline silicon (c-Si),  introducing severe loss from surface amorphization and structural defects. Absorbed two-photon energy density is non-uniformly distributed throughout the cross-section of a waveguide \cite{warner2025simulation}. Combined with tight optical mode confinement, this non-uniformly elevates local peak intensities and accelerates non-linear absorption beyond predictions based on average incident power, creating unevenly distributed amorphous silicon (a-Si) regions and lattice dislocations. Optical modes within the waveguide interact with these disordered regions, inducing significant linear absorption loss. Secondary loss mechanisms include incubated nuclei that behave as Rayleigh scattering centers, increasing round-trip loss following nano-milling threshold over medium pulse repetition counts \cite{zhang2024precision}. Finally, localized stress and micro-fracturing can occur from the ultra-fast energy deposition when laser-induced acoustic shockwaves exceed the fracture toughness of the crystalline silicon, producing micro-cracks along crystallographic planes. At high pulse counts, the incubation effect lowers the silicon damage threshold until the crystal structure  physically collapses  \cite{zhang2024precision,beranek2023melting}.

Alternative approaches include ion implantation with post-annealing \cite{oosterbeek2016effects}, strain engineering via stressor surface adsorption layers \cite{schriever2012strained,callahan_simple_2020}, photon-induced trimming of chalcogenide-assisted silicon waveguides \cite{canciamilla2012photo}, and laser trimming of the operating wavelength of silicon nitride racetrack resonators \cite{de2020laser,kalashnikov2022tuning}. To avoid amorphization-induced losses associated with direct silicon core irradiation, commercial trimming platforms (such as those by Femtum Inc., Québec City, Canada) utilize indirect cladding modification via stress-induced elasto-optic index modulation. Targeting the fused silica (\(\text{SiO}_{2}\)) cladding layer surrounding the waveguide with sub-ablation threshold pulses \cite{wu2005femtosecond,zoubir2006laser}. this process avoids material voiding or surface fracturing, inducing a controlled, localized stress field that permanently tunes the embedded waveguide's refractive index.  Since the laser focus remains isolated within the cladding, the crystalline silicon core is spared from pulse-by-pulse nanocrystalline nucleation and amorphization inherent to direct-core femtosecond approaches. Bypassing incubation-driven amorphization also prevents Rayleigh scattering losses, enabling precision correction of manufacturing variations with zero static power consumption, zero cross-talk, and ultra-low insertion loss (<0.1 \si{dB}).

\section{Design Purpose}
The purpose of this passive silicon photonics design is to realize filters and analog signal operators from silicon (Si) and silicon nitride (SiNx) racetrack resonator/modulators. Unique design requirements are satisfied in one-dimensional (1D) analytical,  two-dimensional (2.5D) ANSYS\textsuperscript{\textregistered} Lumerical variable finite-difference time-domain (varFDTD), and fully three-dimensional (3D)  finite-difference time-domain (FDTD) numerical solutions. Unique design requirements are intended to satisfy local measurements of refractive index. ANSYS Lumerical licenses were provided by CMC during the silicon passives (SiEPIC) workshop.

\section{Design Layout Description}
The first four black box edge couplers on the North edge (from left to right) are outputs for three-stage ring resonator/modulators, whereas the fifth black box edge coupler is intended as an input to the ring resonator/modulator $4\times4$ matrix multiplier, and the remaining sixth black box edge coupler is intended as the output of the ring resonator/modulator $4\times4$ matrix multiplier. Waveguides from the four left-most black box edge couplers receive output signals from two third-order ring resonators/modulators (2 edge couplers for through ports and 2 edge couplers for drop ports). Each third order ring resonator/modulator consists of a first stage ring coupled to a linear input waveguide connected from a South black box  edge coupler to a North  black box edge coupler as the through port, and a third-stage ring couples an output waveguide that is terminated at one end, and has its drop port curved 180$^\circ$ toward a North edge coupler black box cell. Each silicon edge coupler on the South edge of the floorplan is a standard black box cell with only one input or output branch. The first two black box edge couplers (from left to right) are inputs to the third order ring resonator/modulators. The remaining four black box edge couplers on the South end of the floorplan transmit outputs from a 4$\times$4 ring resonator/modulator matrix multiplier. In the 4$\times$4 ring resonator/modulator matrix multiplier, one Y-branch from a North black box edge coupler splits the input into two additional Y-branches that are input to linear waveguides that intersect at  the North waveguide crossing of the $4\times4$ matrix multiplier. The remaining North black box edge coupler receives output from the reverse drop port of each resonator/modulator in the array (for debugging purposes).  Near the South West corner of each waveguide crossing is a ring resonator/modulator that couples to the nearest opposite waveguide which transmits the resonant output as the drop port. Each output from the $4\times4$ matrix multiplier is transmitted into independent  black box edge couplers on the South end of the floorplan. All remaining through ports are terminated. Last are two silicon nitride (SiNx) ring resonator/modulators with rib-waveguide coupling. Each  ring resonator/modulator is coupled to the same two waveguides, which are also connected to specialized edge couplers with 65 \si{\micro m} optimized terminated tapers of SiNx ending near the edge of the floor plan (\(\sim\)1 \si{\micro m} away). The tapers are designed for a specific field envelope profile matching lensed optical fibers of a conventional test setup. The two SiNx rings have a very different radius, resulting in a different order of magnitude and intensity output signal, which are assumed to be distinguishable in testing. 
\vspace{-16mm}
\begin{figure}[H]
    \centering
    \includegraphics[width=\linewidth]{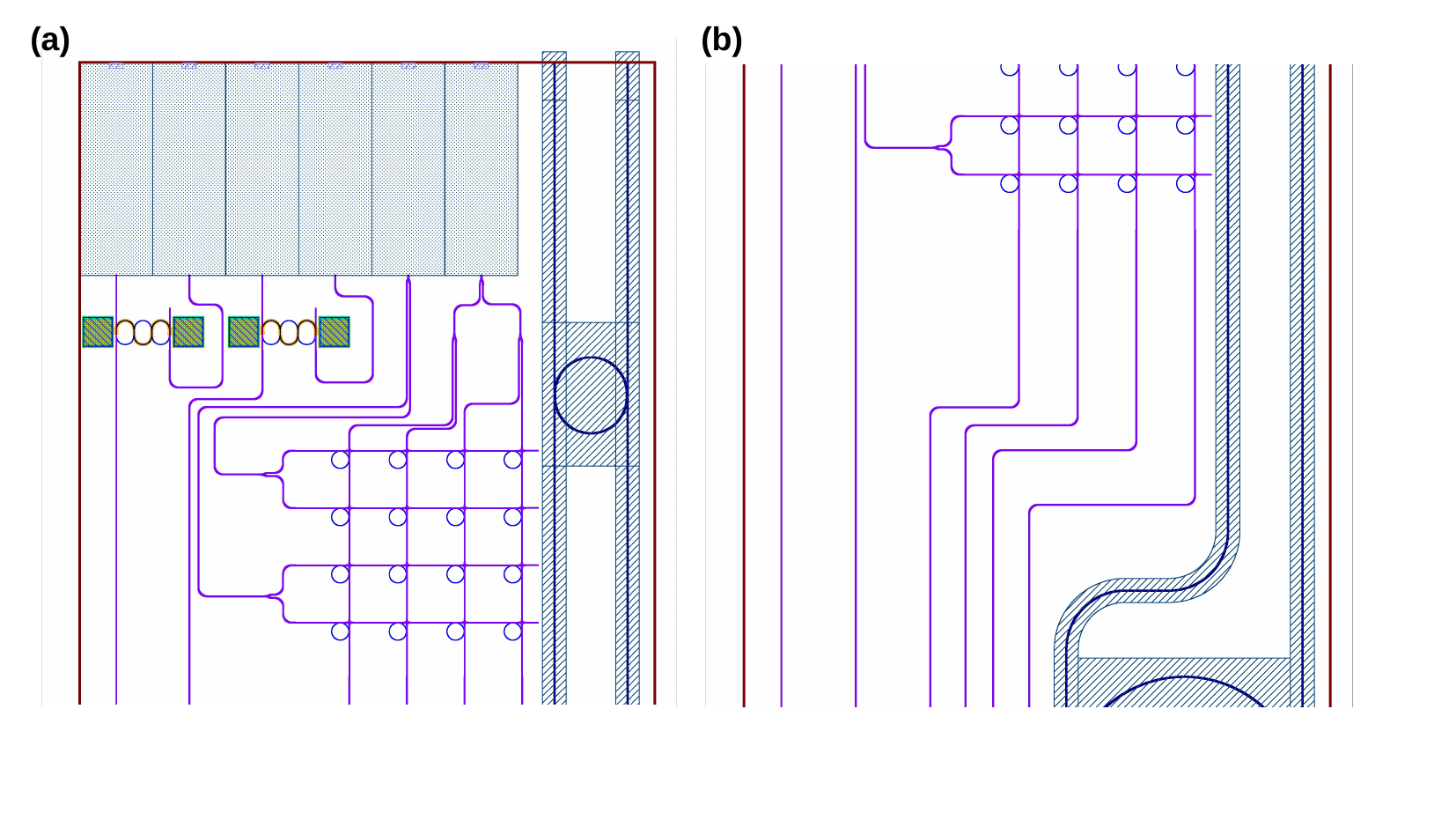}\vspace{-8mm}
    \includegraphics[width=\linewidth]{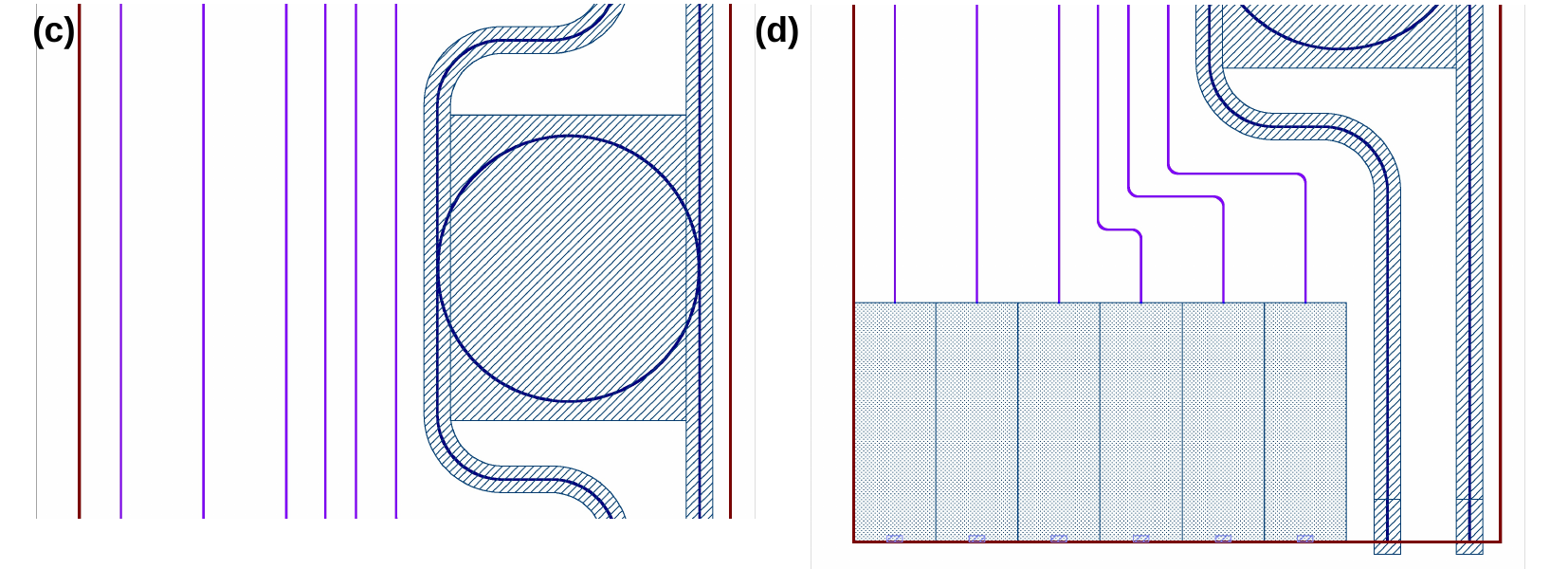}\vspace{-8mm}
    \includegraphics[width=\linewidth]{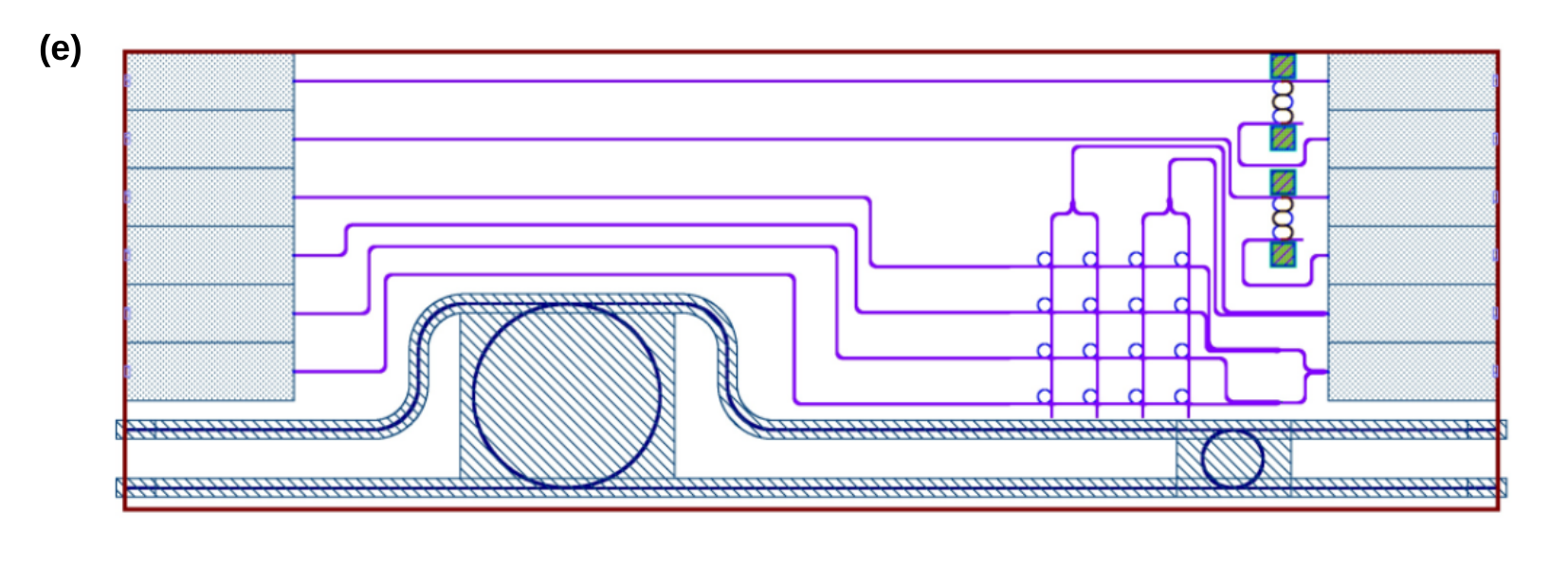}\vspace{-8mm}
    \caption{Full hierarchy of layout in Klayout software. Black box edge-couplers interface from the bottom and top edges of the silicon wafer.}
    \label{fig:1}
\end{figure}

\section{Design purpose for layout cells}\label{sec:design_purpose}
In this section, the purpose of each cell in the layout is described in further detail.

\subsection{Third order racetrack resonator/modulator}
The third order racetrack resonator/modulator outputs a broadband through port and narrowband drop port to realize a Butterworth filter. There is a 50$\times$50 \si{\micro m} wire bond contact (metal contact layer) which is connected via an integrated micro-heater layer heating pad used to modify the effective index along the waveguide. Modifying effective index changes the resonant frequency of each ring simultaneously, which can be measured as a wavelength shift in the resonance spectrum.

\begin{figure}[H]
    \centering
    \includegraphics[width=\linewidth]{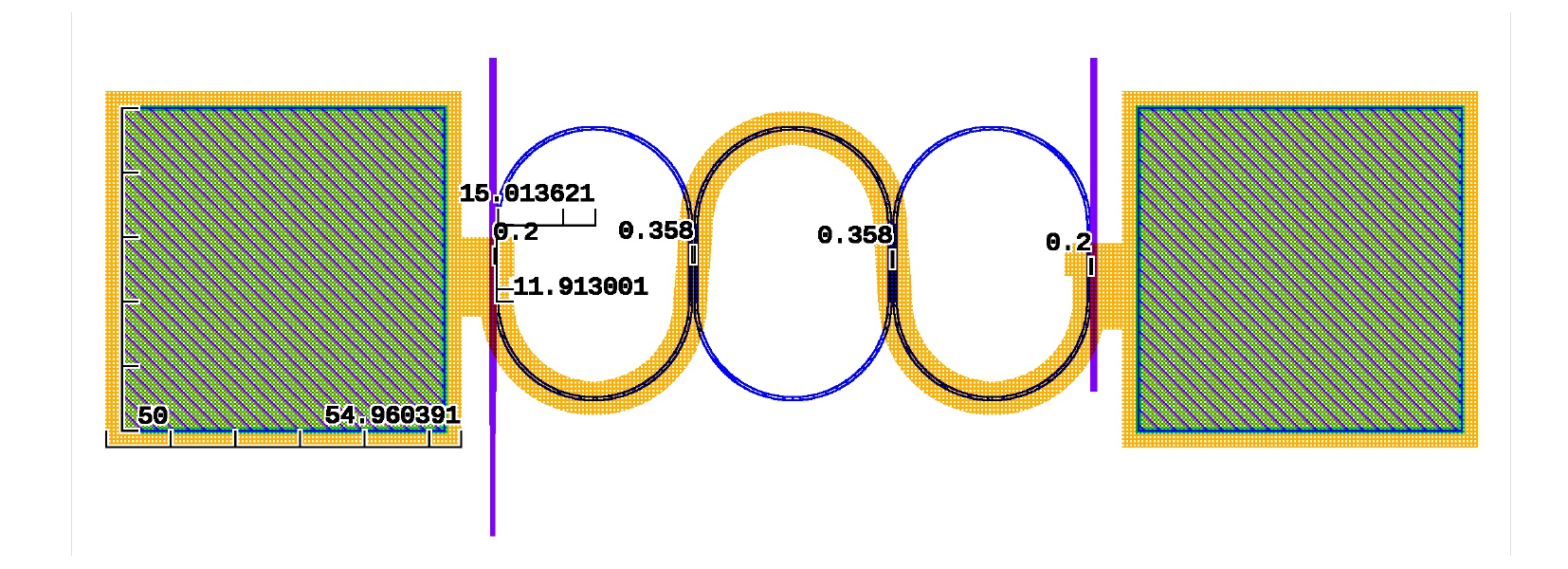}
    \caption{Full hierarchy of the third order resonator/modulator. Thermal contact wirebonds (metal contacts layer) are 50 \si{\micro m} along each side length and are surrounded by 55 \si{\micro m} thermal pads (integrated micro-heaters layer) that interlace the ring resonator/modulator in an S-pattern with a 5 \si{\micro m} width. Each of the racetrack resonator/modulators has a 15 \si{\micro m} radius and a coupling length of 11.913 \si{\micro m}. Each linear coupling length nearest the through port or drop port waveguides shares a gap of 200 \si{nm}, whereas the center resonator shares a 358 \si{nm} gap from each adjacent resonator/modulator. The response of the third-order resonator/modulator is a Butterworth filter with a detune frequency bandwidth of $\Delta\nu_B=50$ \si{GHz} and a free spectral range (FSR) of 5.3 \si{nm}. }
    \label{fig:2}
\end{figure}

\subsection{Ring resonator/modulator matrix multiplier}
The ring resonator/modulator 4$\times$4 matrix multiplier provides output resonance signals as a linear combination of input signals, with coefficients being the transfer functions of each resonator/modulator. Located in the South East corner of each \texttt{Microring\_array\_element} cell is a ring resonator/modulator of 15 \si{\micro m} radius which shares a 150 \si{nm} coupling gap to two linear waveguides (each intersecting the waveguide crossings at 90$^\circ$) . The matrix multiplier is realized as a two-dimensional (2D) $4\times4$ periodic array of the \texttt{Microring\_array\_element} cell as the \texttt{Microring\_array} cell. Therefore, the \texttt{Microring\_array} cell exhibits a matrix transfer function, and each element of the matrix is the transfer function of the \texttt{Microring\_array\_element} cell. Therefore, each output of the \texttt{Microring\_array} cell is a linear combination of its inputs. In the ideal case of exact similar ring resonators/modulators, and equivalent inputs, the transfer function of the \texttt{Microring\_array} cell reduces to the transfer function of the \texttt{Microring\_array\_element} cell. However, process variations from the ideal transfer function and spectrum are expected to occur. Changes to refractive index of each element in the array can occur from femtosecond ablation, or prolonged exposure to sufficiently intense light from above the waveguide, resulting in increased two-photon absorption which modifies the crystal structure. Idealized performance for telecommunications would necessitate the use of active heaters to achieve specific resonance criteria given defects and error tolerances in fabrication. However, for purposes of a diagnostic, these heaters are not required if the device can be measured both before and after ablation.

\begin{figure}[H]
    \centering
    \includegraphics[width=\linewidth]{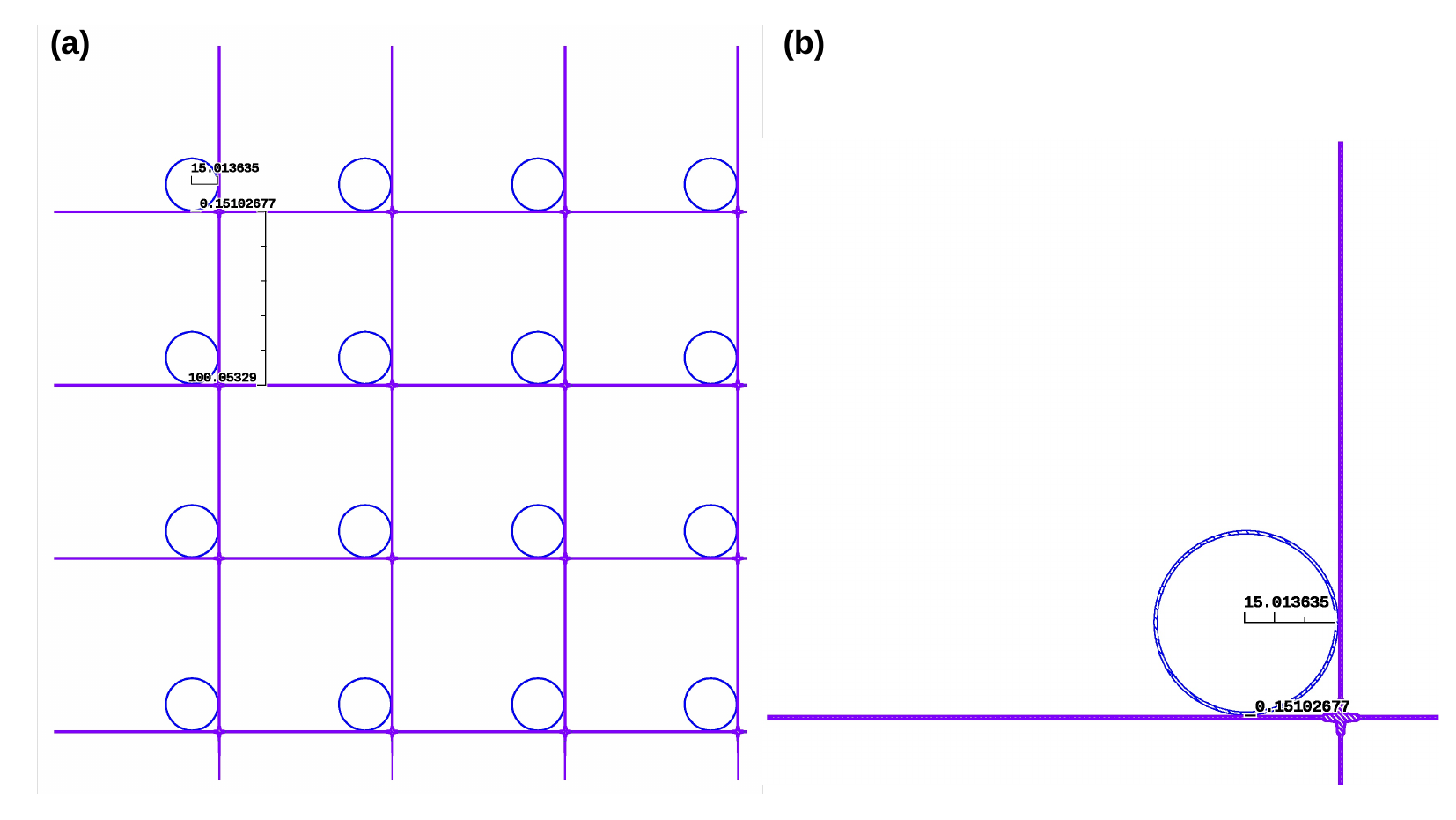}
    \caption{Full hierarchy of ring resonator/modulator $4\times4$ matrix multiplier in KLayout (cell named \texttt{Microring\_array}). Each ring has a 15 \si{\micro m} radius and shares a 150 \si{nm} gap from the nearest linear waveguide. Input ports are along the top, whereas the through ports are directed south, and the drop-ports are directed left. The ports opposite to the drop port are used as secondary inputs for debugging purposes. The waveguide length between adjacent array elements is \(\sim\)90 \si{\micro m}, and each waveguide intersects waveguide crossings. }
    \label{fig:3}
\end{figure}
\subsection{Silicon nitride (SiNx) racetrack resonators/modulators}
There are two different silicon nitride (SiNx) racetrack resonator/modulators that are coupled in parallel with the same two SiNx rib waveguides with 1 \si{\micro m} width.
\begin{itemize}
    \item \textbf{Resonator/modulator 1}: Racetrack with a \(\sim\)65\si{\micro m} radius, $L_c=8$\si{\micro m} coupling length, and 300 \si{nm} gap to the nearest rib waveguide.
    \item \textbf{Resonator/modulator 2}: Racetrack with a \(\sim\)200\si{\micro m} radius, $L_c=8$\si{\micro m} coupling length, and a 320 \si{nm} gap to the nearest rib waveguide.
\end{itemize}
The purpose of this device is to measure the effective index variation and loss from the SiNx process for comparison to the silicon racetrack and ring resonators/modulators. In addition, the amorphous structure of the SiNx ring resonators can be used to track the influence that optical polarization has with respect to the anisotropic silicon structure.
\begin{figure}[H]
    \centering
    \includegraphics[width=\linewidth]{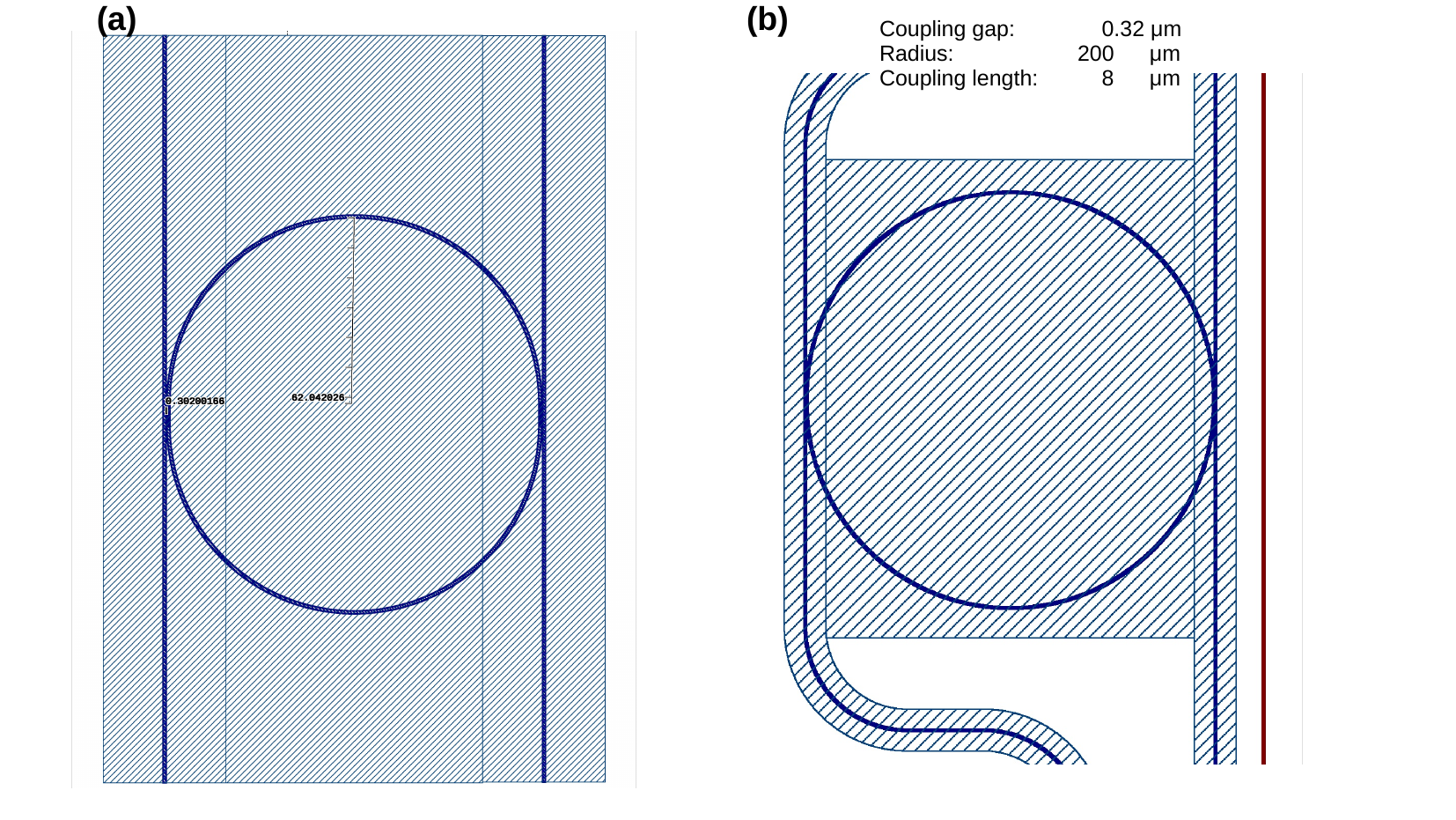}
    \caption{Silicon nitride (SiNx) racetrack resonators in Klayout. (a) \( \sim \)65 \si{\micro m} radius racetrack resonator and (b) \(\sim\)200 \si{\micro m} radius racetrack resonator. Each SiNx resonator has a rib waveguide structure with a 20 \si{\micro m} spacing.}
    \label{fig:4}
\end{figure}

\subsection{Silicon nitride (SiNx) taper}
There is an SiNx taper of 65 \si{\micro m} length starting from a 1.1 \si{\micro m} width and ending in a 1.0 \si{\micro m} width to satisfy the standard guided mode in the SiNx rib waveguide. The cell is labeled \texttt{SiNx\_edge\_coupler} and overlaps the North and South edges of the floorplan in the deeptrench to satisfy the 20 \si{\micro m} spacing of silicon nitride trench window. One critical dimension is the 1 \si{micro m} spacing of the edge coupler from the floorplan.
 
\section{Expected performance}
Each device is expected to perform differently based on the function described in the ``Design purpose for layout cells” (section \ref{sec:design_purpose}). Analytical models of the output spectra are provided only for the third-order ring resonator/modulator, and a single element of the $4\times4$ matrix multiplier, which are compared directly with the simulated output spectra in ANSYS\textsuperscript{\textregistered}  Lumerical FDTD, and ANSYS\textsuperscript{\textregistered}  Lumerical varFDTD. Since we have limited knowledge of the AMF process variation, it is not characterized in detail.

\subsection{Third-order coupled resonator/modulator} 
The third-order ring resonator/modulator is modeled using an estimation of the transmitted signal from the coupling coefficient given the expected \mbox{-}3 \si{dB} bandwidth of the input signal, the radius of the ring, and the group index assuming a linearized dispersion relation. Given the designed 15 \si{\micro m} radius, 11.913 \si{\micro m} coupling length, and  $\Delta\nu_B=50$ \si{GHz} detuned frequency bandwidth, the output resonant peak is estimated to occur at 10 \si{dBm} with a free spectral range of 5.3 \si{nm}. These expected performance parameters are listed in Table \ref{tab:1}, and were derived from analytic models provided by Robin Kim.

\begin{table}[H]
    \centering
    \caption{Three-stage ring resonator performance metrics}
    \begin{tabularx}{\textwidth}{|X|c|c|}
    \hline
        Parameter & Value & Units \\
        \hline \hline
        Free spectral range ($\lambda = 1550\text{nm}$) & 5.3 & \si{nm} \\
        \hline
        Bandwidth & 50 & \si{GHz} \\
        \hline
        Transmitted SNR & 10 & \si{dBm} \\
        \hline
    \end{tabularx}
    \label{tab:1}
\end{table}

Figure \ref{fig:5} shows an annotated diagram of the third-order resonator/modulator analytical model and the expected spectrum given ideal fabrication. Since the resonators are in series, the same resonance should occur. However, due to machine tolerance and process variation, it is expected that the output signal will be weaker due to constructive interference between the rings.

\begin{figure}[H]
    \centering    \includegraphics[width=0.49\linewidth]{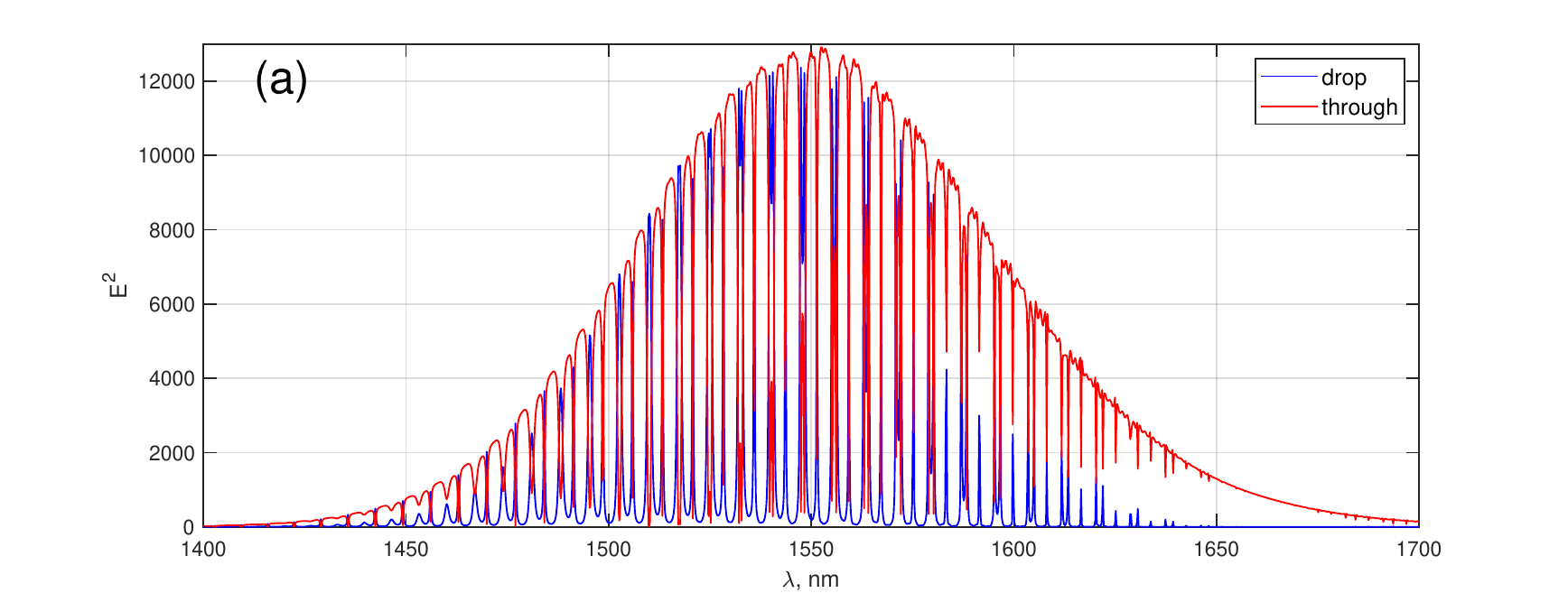}
    \includegraphics[width=0.49\linewidth]{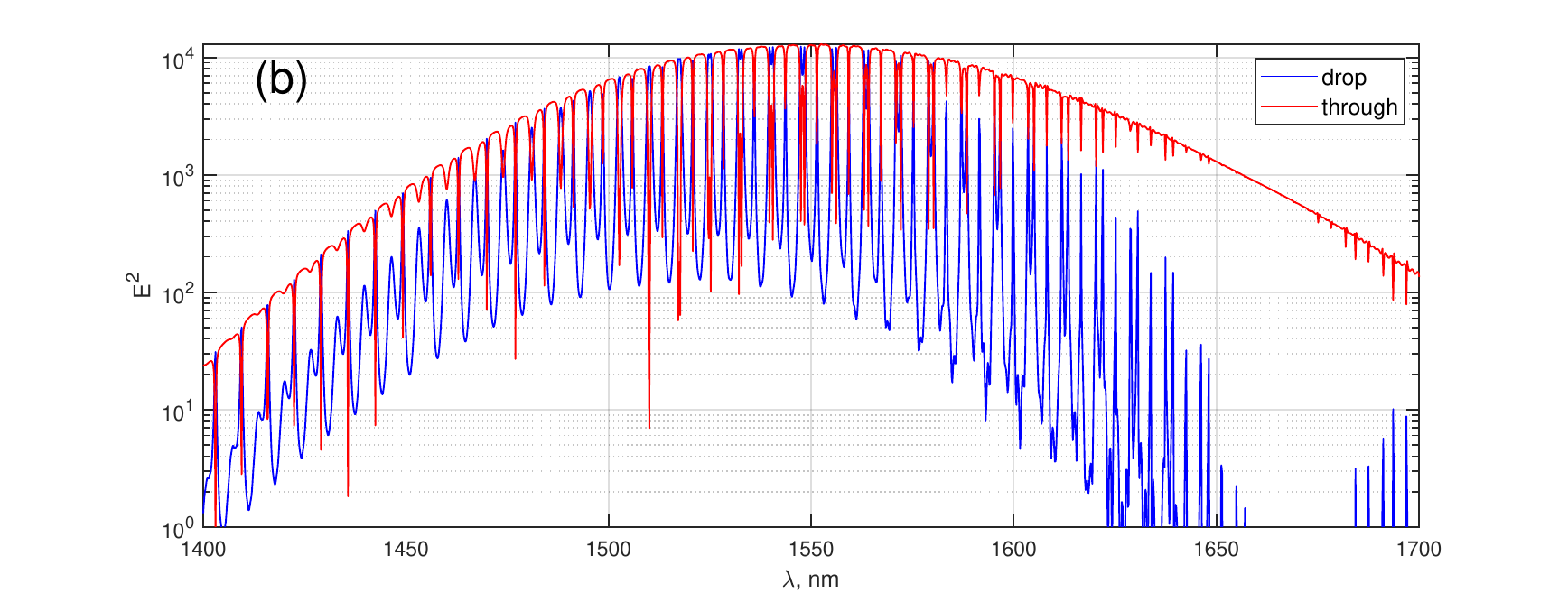}
\includegraphics[width=0.49\linewidth]{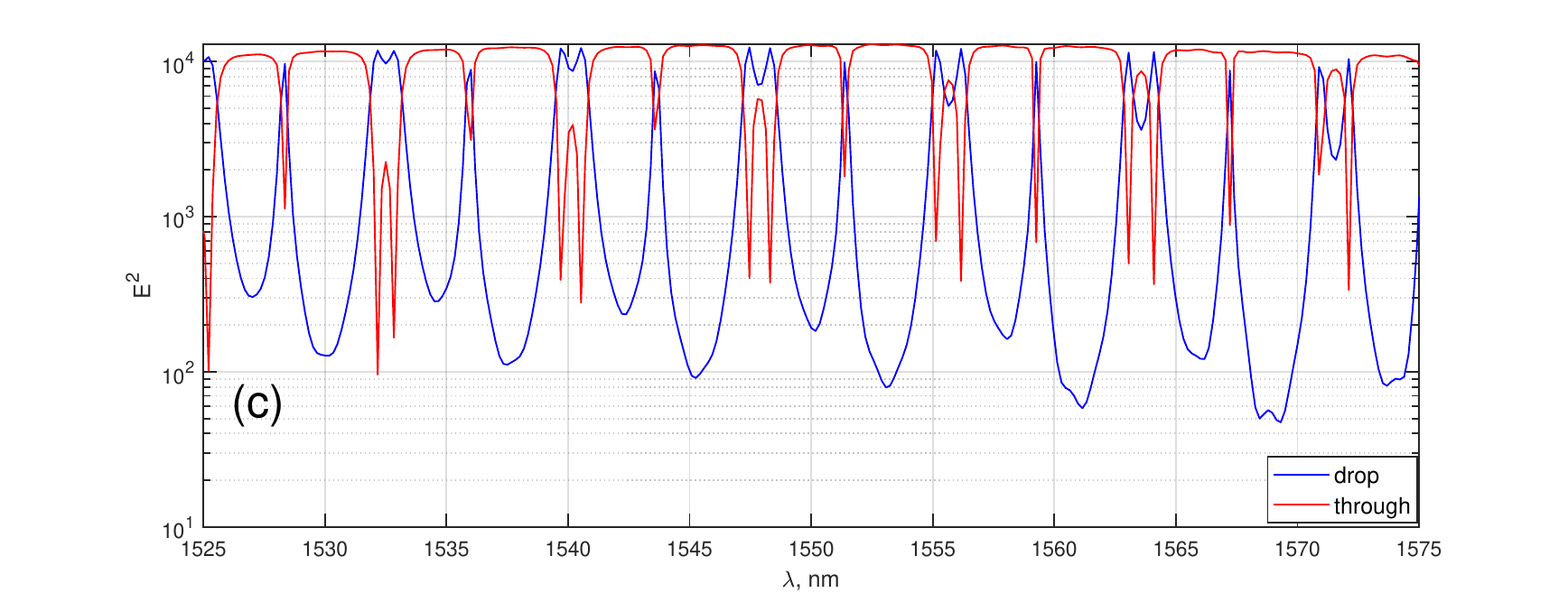}
\includegraphics[width=0.49\linewidth]{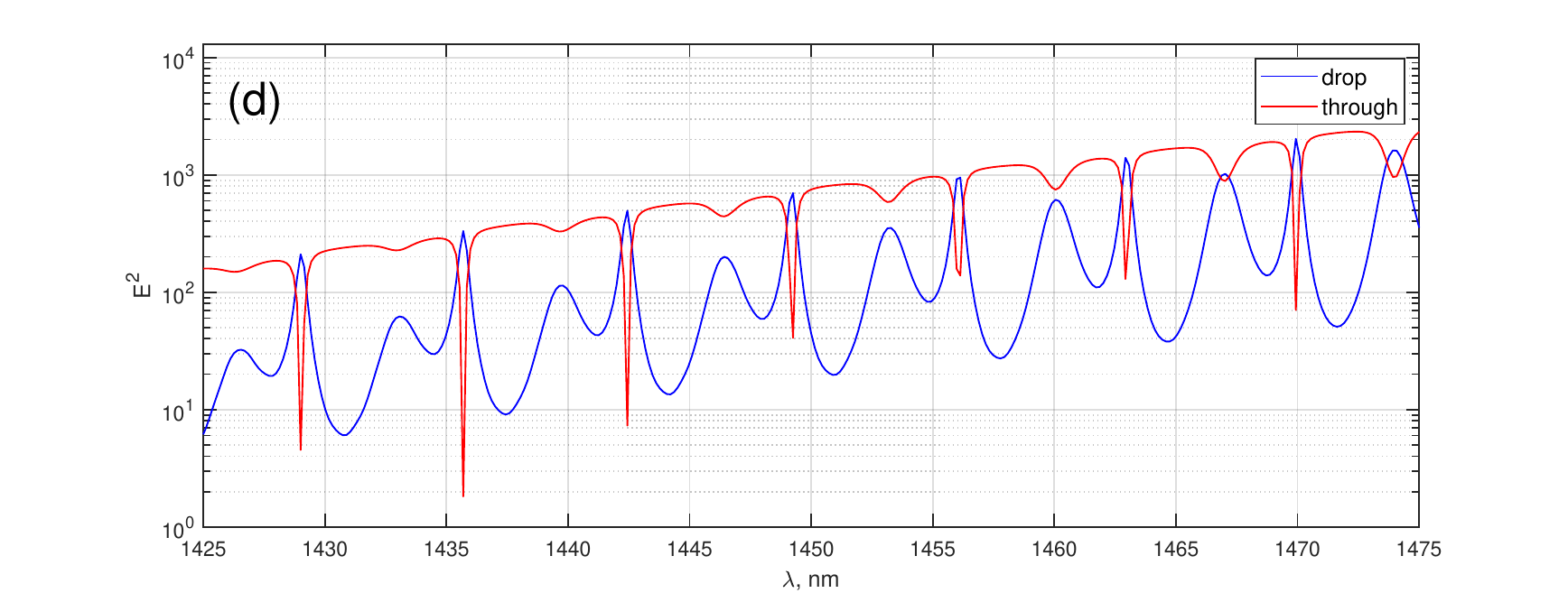}
    \caption{The wavelength spectra of the third-order racetrack resonator/modulator after 40000 \si{fs}. }
    \label{fig:7}
\end{figure}

Fig. \ref{fig:6} shows an ideal filter response, but the relative spectral response is differs from propagation and modal dispersion, which are not accounted for in the simplified analytical model. In order to characterize the response of the third-order ring resonator/modulator, ANSYS\textsuperscript{\textregistered} Lumerical varFDTD is used to model the temporal signal in the center of the waveguide at the through and drop ports. Spectra from each time-domain signal reveal both broadband and narrowband responses atypical of the simplified analytical third-order racetrack resonator/modulator model, as shown in Fig. \ref{fig:7}.

\subsection{Ring resonator/modulator matrix multiplier}
\begin{figure}[H]
    \centering
    \includegraphics[width=0.49\linewidth]{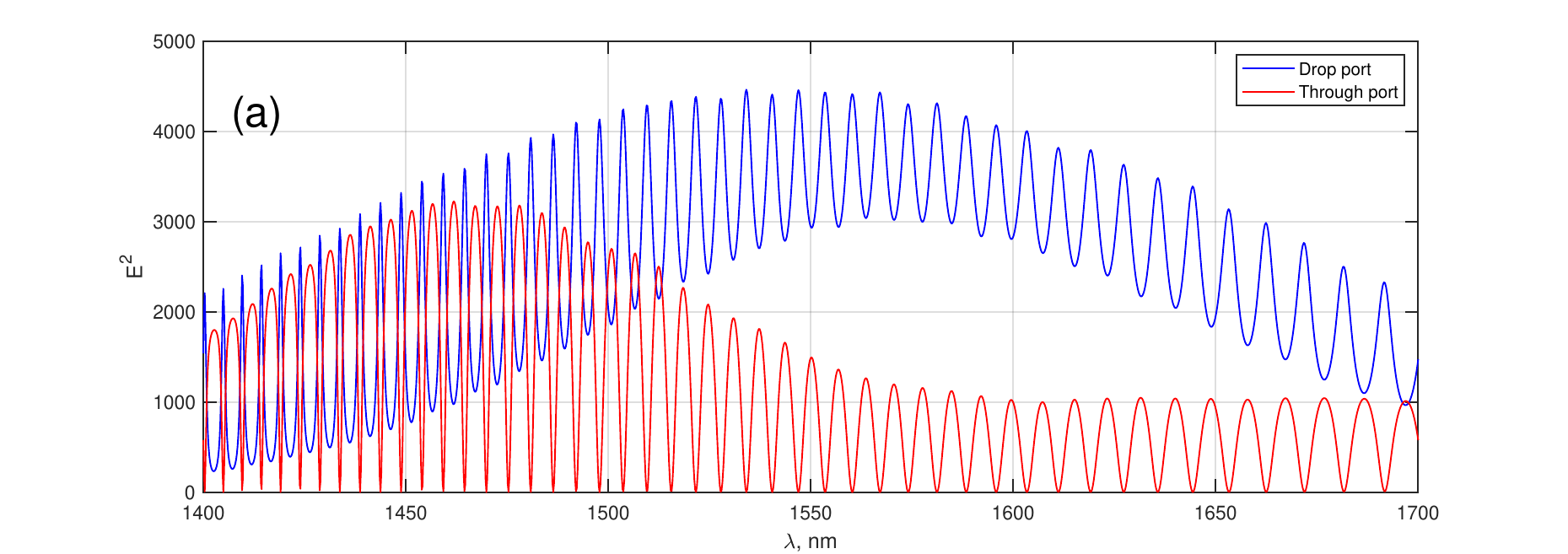}
    \includegraphics[width=0.49\linewidth]{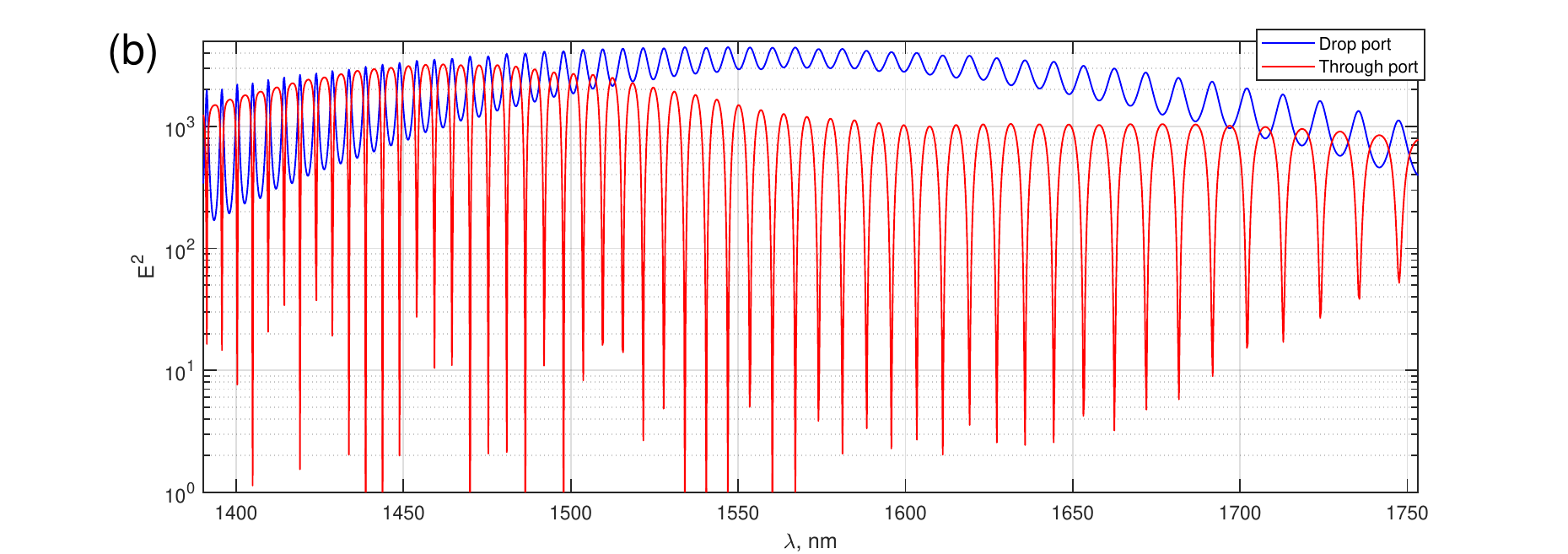}

    \includegraphics[width=0.49\linewidth]{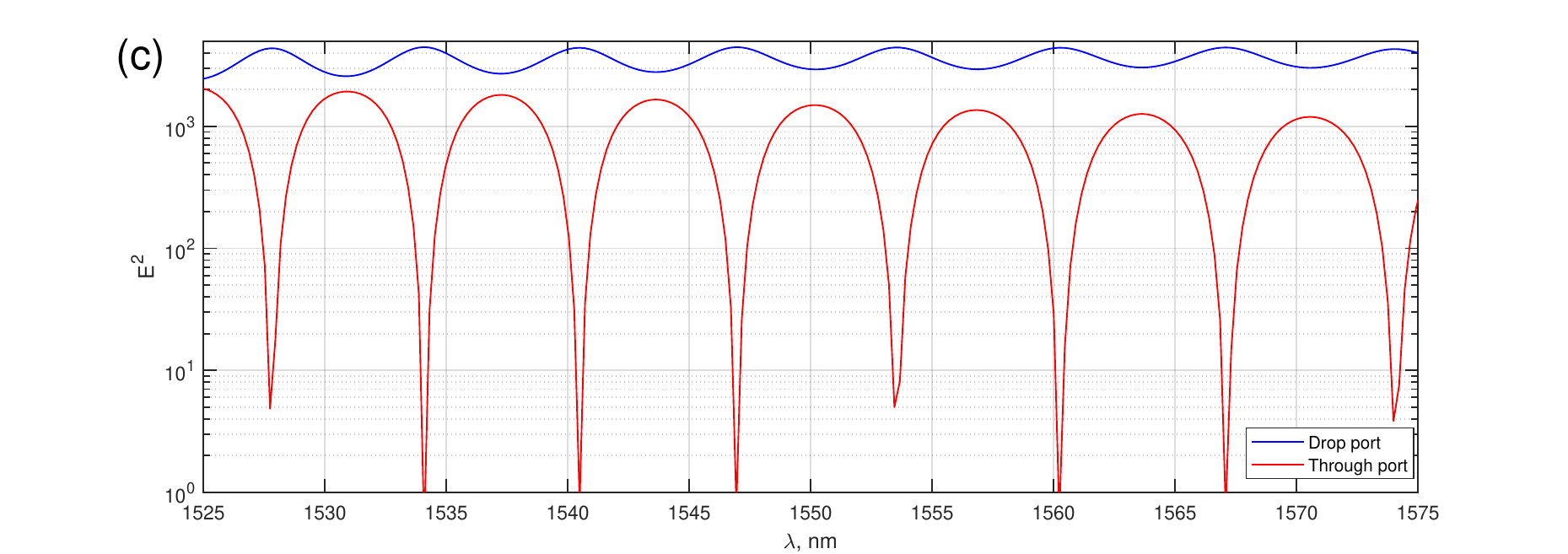}
        \includegraphics[width=0.49\linewidth]{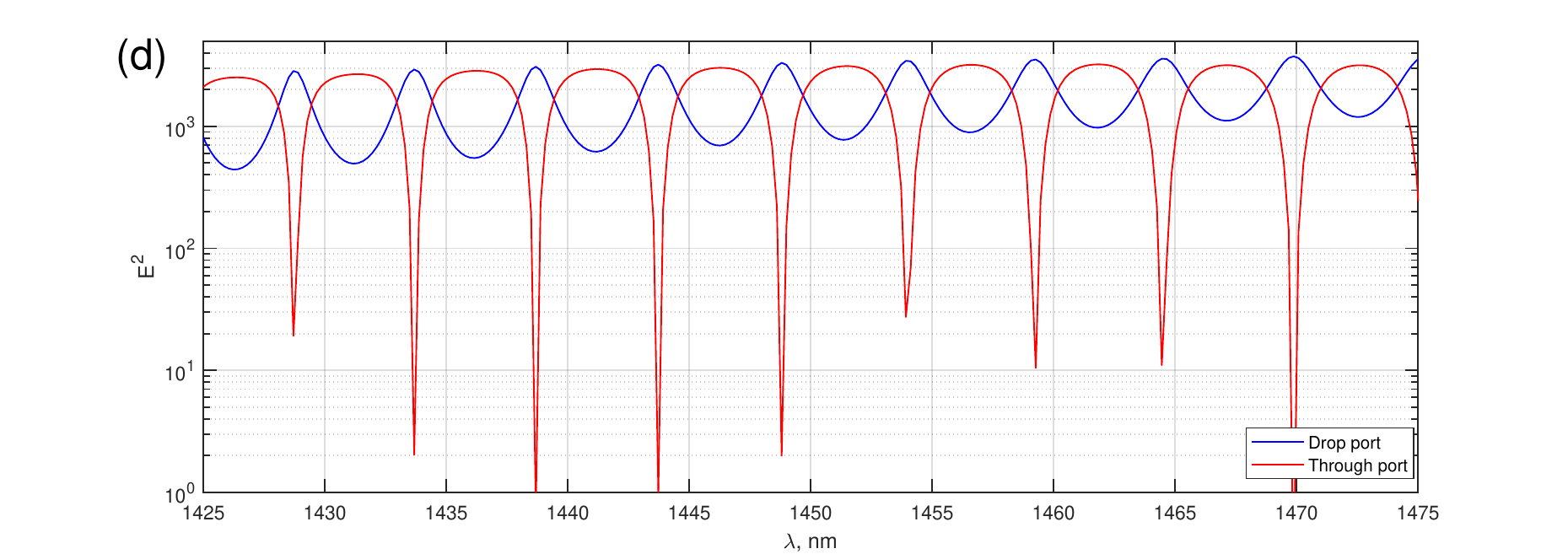}
    \caption{The relative intensity of the through and drop port spectra for (a) $\lambda\in[1400,1700]$ \si{nm}; (b) same as (a) on a $\log_{10}$ scale; (c) $\lambda\in[1525,1575]$ \si{nm}; and (d) $\lambda\in[1425,1475]$ \si{nm}. Since the power coupling model would account for a phase shift due to the unique path length from the case of a two-waveguide resonator, the spectrum has a weaker resonance at the expected wavelength, and a stronger resonance at a shorter wavelength.}
    \label{fig:9}
\end{figure}

Performance metrics for the ring resonator/modulator $4\times4$ matrix multiplier are based on measurements of one previously fabricated AMF ring resonator/modulator. The single input signal is transmitted into multiple outputs, and any variation from one of the matrix elements will enter as modulation/variation in the linear sum amplitude output spectra. The simulated resonance spectrum for one of the rings is provided in Fig. \ref{fig:9}.

The equivalent simulated optical spectrum is analyzed using ANSYS\textsuperscript{\textregistered} Lumerical varFDTD. It exhibits disagreements with the amplitude of the analytical spectrum given an energy coupling model, most likely as a consequence of reduced optical path length from the 90$^\circ$ arc as opposed to a 180$^\circ$ arc. These differences may be accounted for using a power coupling model, however, the simulated spectrum demonstrates the bandwidth and FSR are unaffected. Therefore, the device functionality is sufficient.


\begin{figure}[H]
    \centering    \includegraphics[width=\linewidth]{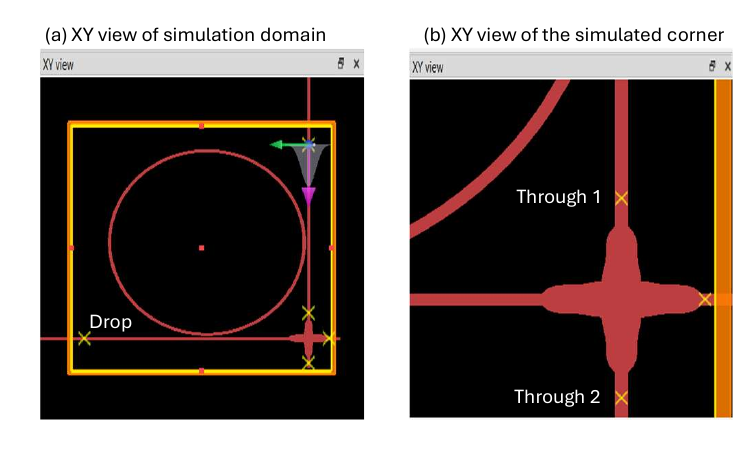}
    \caption{The corner resonator simulation domain in ANSYS\textsuperscript{\textregistered} Lumerical varFDTD. Time monitors are placed at the center of the through and drop ports, and monitor the signal over 20000 \si{fs}.}
    \label{fig:8}
\end{figure}

Ring resonators with zero coupling length are chosen for the array in order that an under-coupled response is achieved with the desired resonance spectrum. The purpose of under-coupling in this case is to ensure that sufficient power reaches consecutive matrix elements. However, the limitation of under-coupling is increased modal dispersion and losses due to the amf\_Crossing, which is unavoidable.
\begin{figure}[H]
    \centering
    \includegraphics[width=\linewidth]{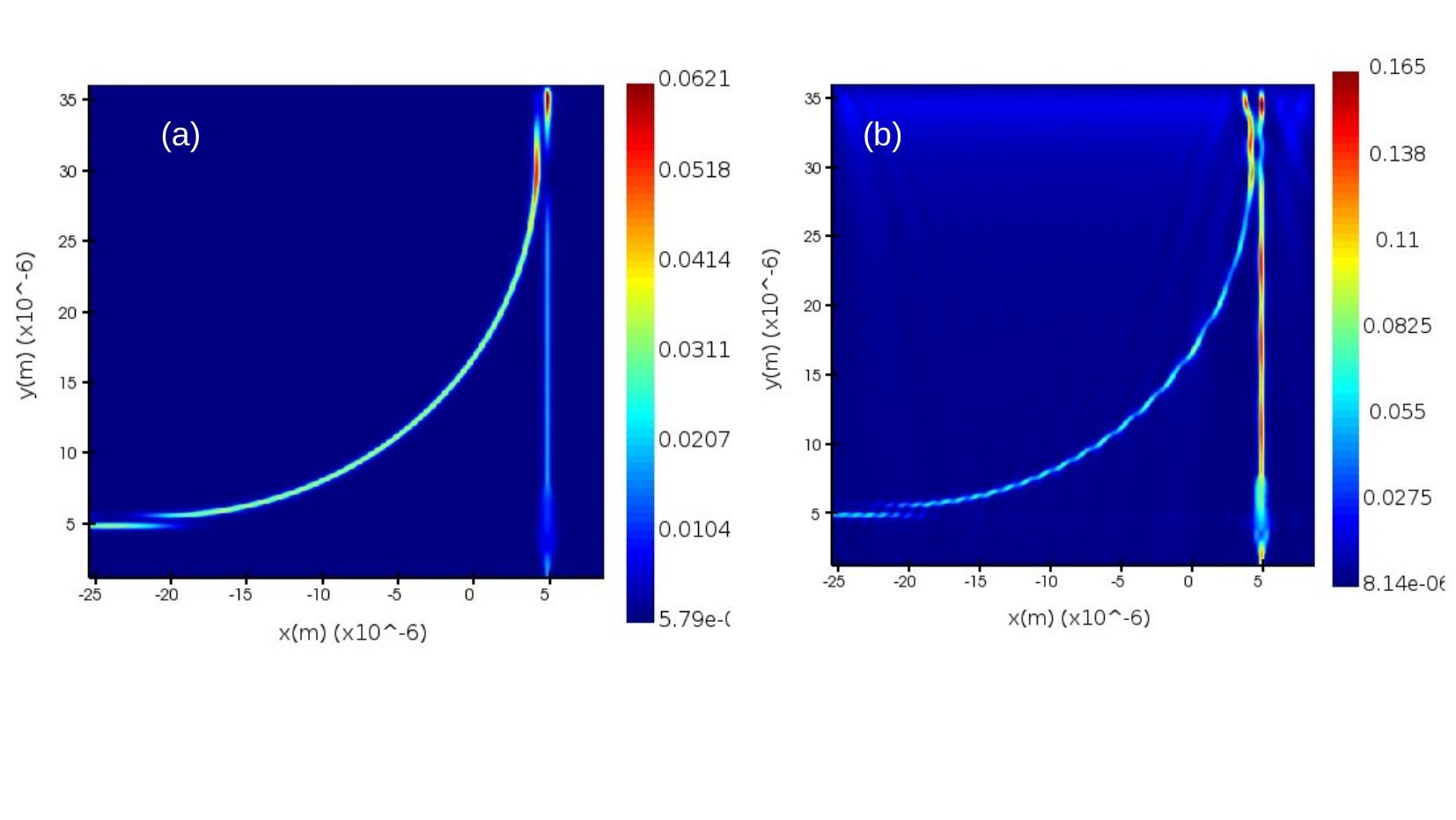}
    \caption{Demonstration of coupling at the corner. (a) The power given a coupling length $L_c=2$ \si{\micro m}. (b) The coupling length $L_c=0$ \si{\micro m} provides greater power for consecutive ring resonators in the array. In this case, balanced coupling is preferred. }
    \label{fig:10}
\end{figure}

\subsection{Silicon nitride (SiNx) ring resonators}
\begin{figure}[H]
    \centering
    \includegraphics[width=0.49\linewidth]{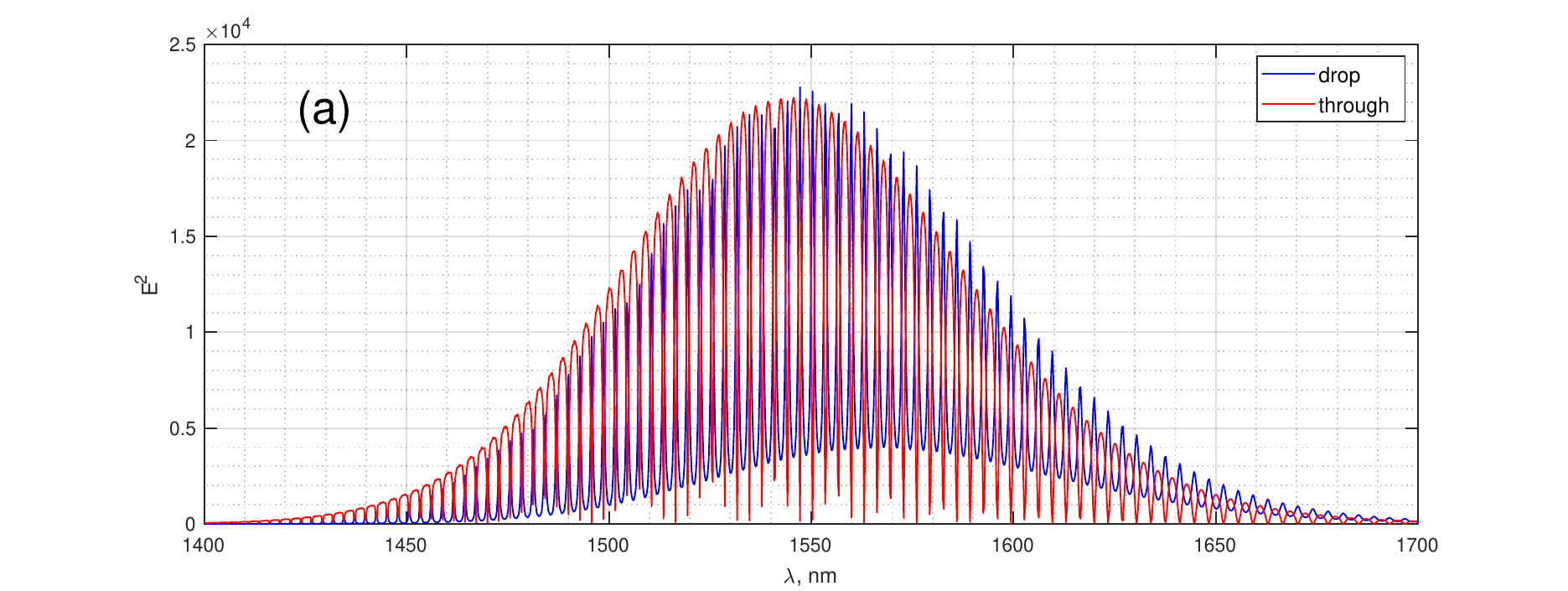}
    \includegraphics[width=0.49\linewidth]{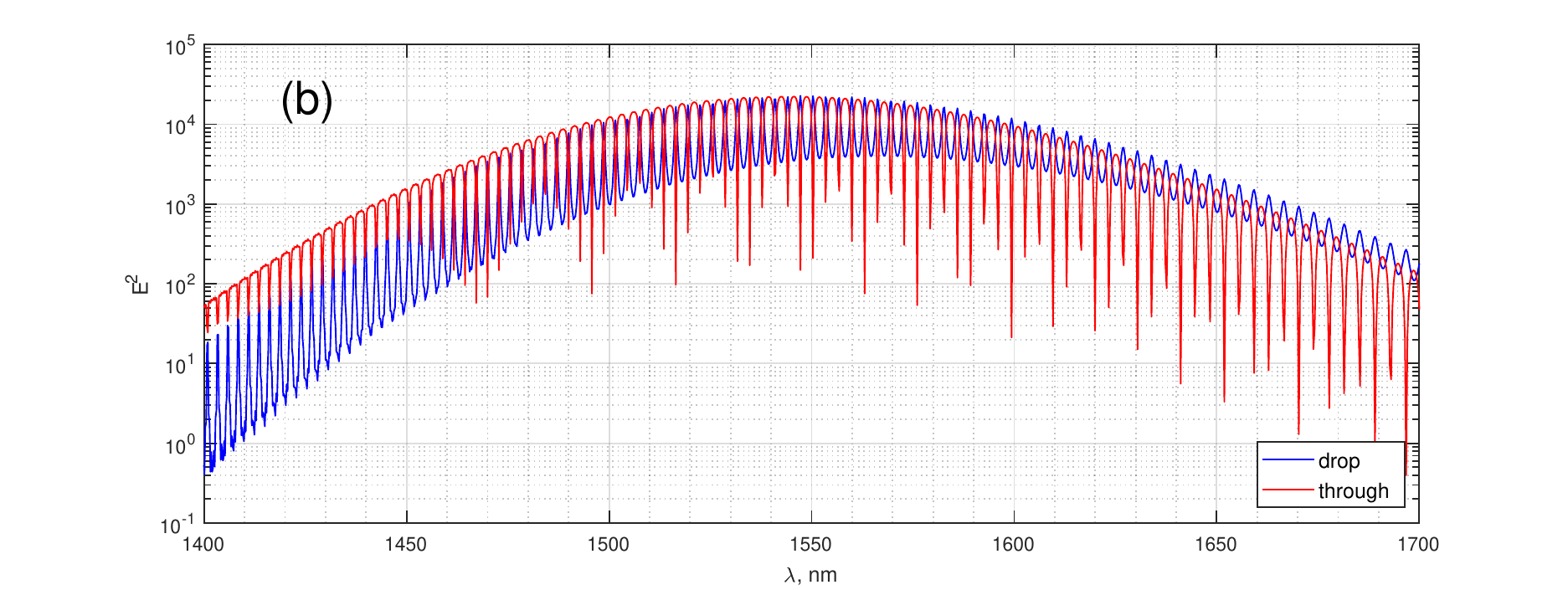}
    \includegraphics[width=0.49\linewidth]{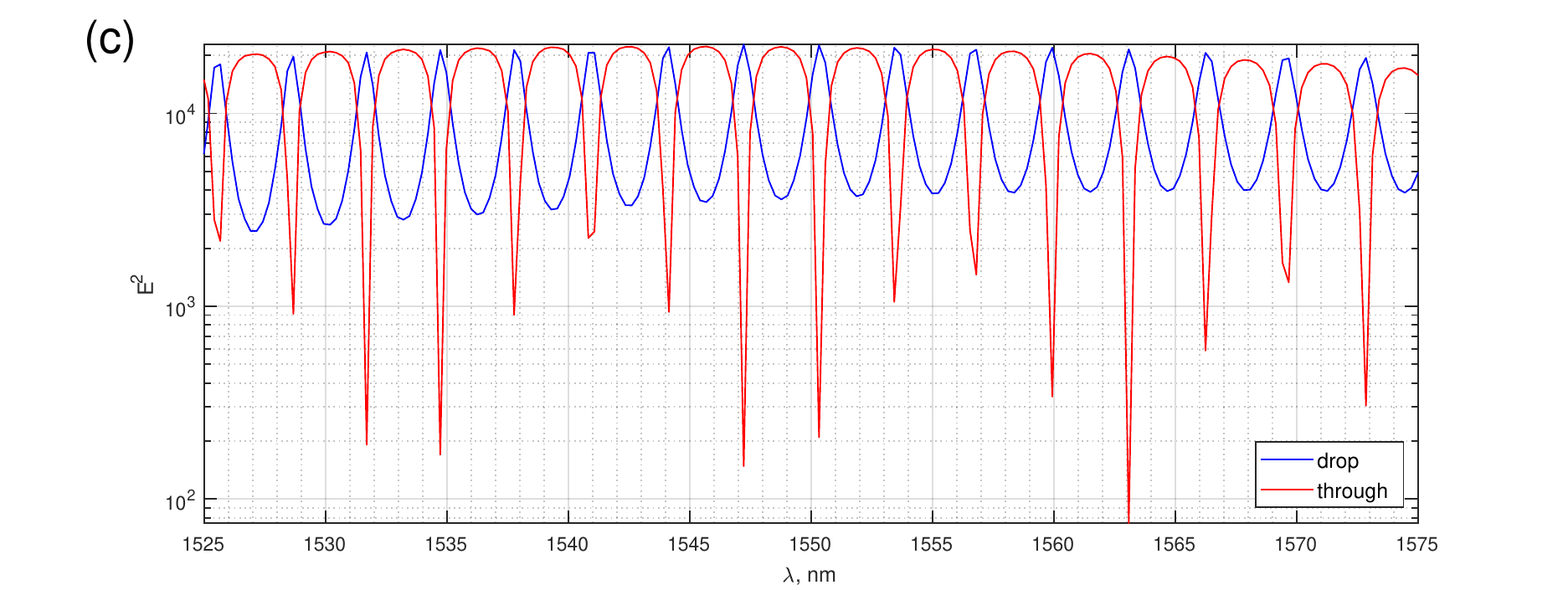}
    \includegraphics[width=0.49\linewidth]{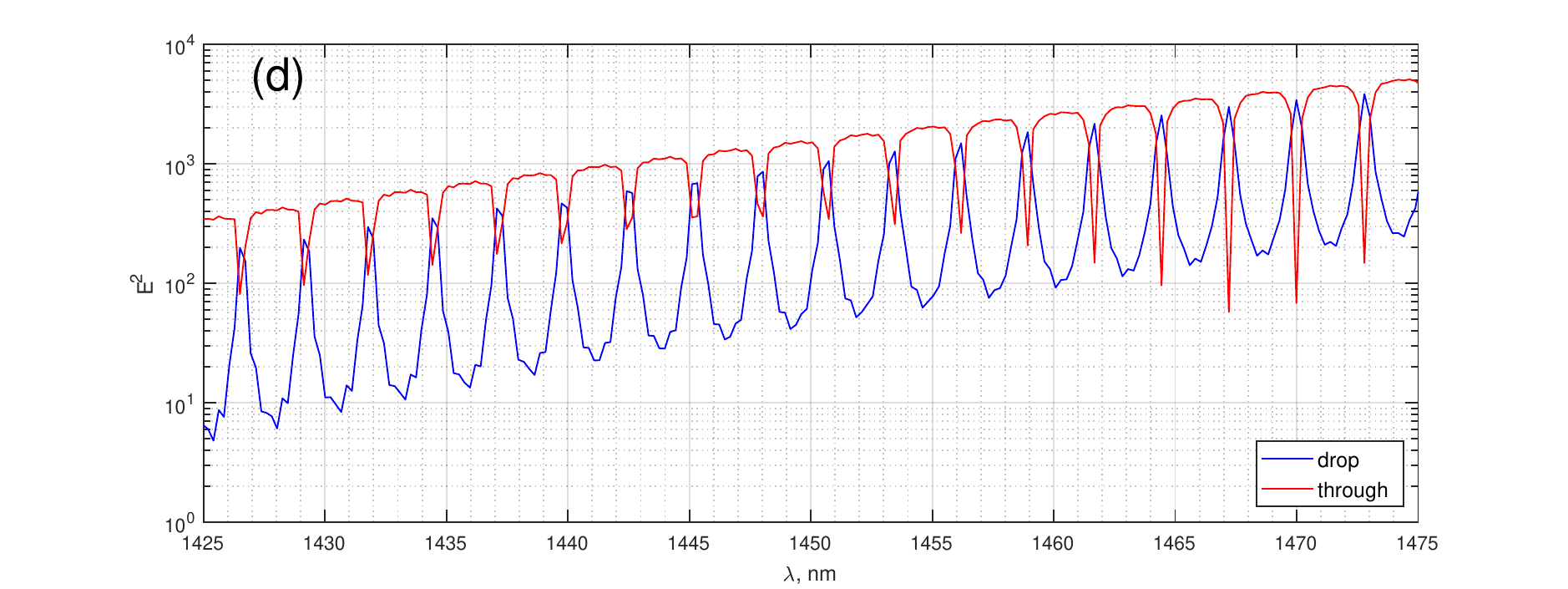}
    
    \caption{Spectra for the \( \sim \) 65 \si{\micro m} radius racetrack resonator/modulator with 8 \si{\micro m} coupling length simulated in ANSYS\textsuperscript{\textregistered} Lumerical varFDTD.}
    \label{fig:11}
\end{figure}

\begin{figure}[H]
    \centering
    \includegraphics[width=0.8\linewidth]{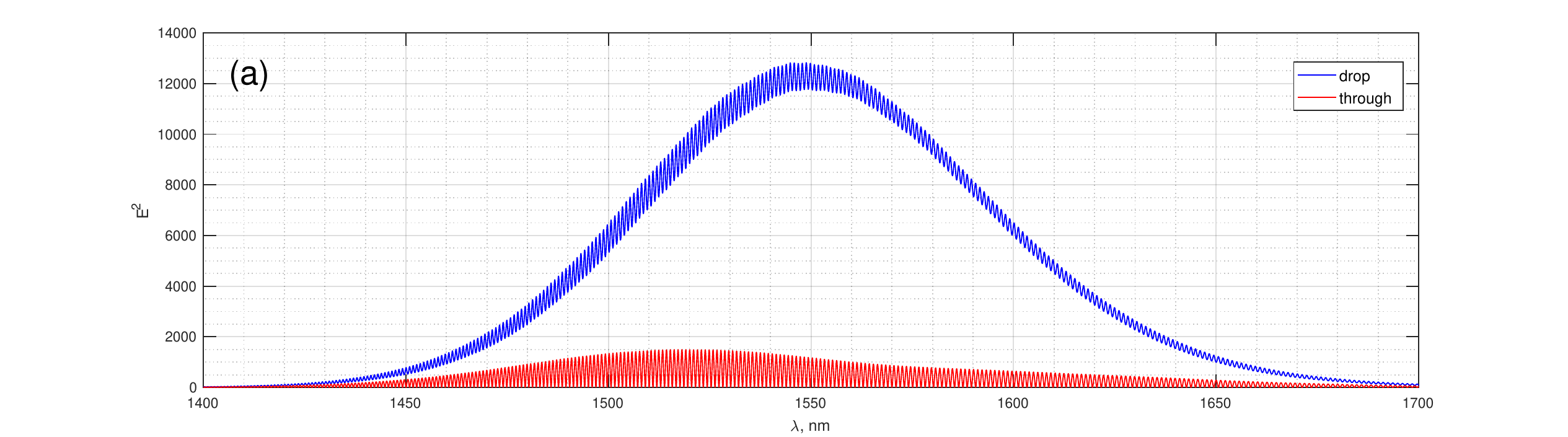}
    \includegraphics[width=0.8\linewidth]{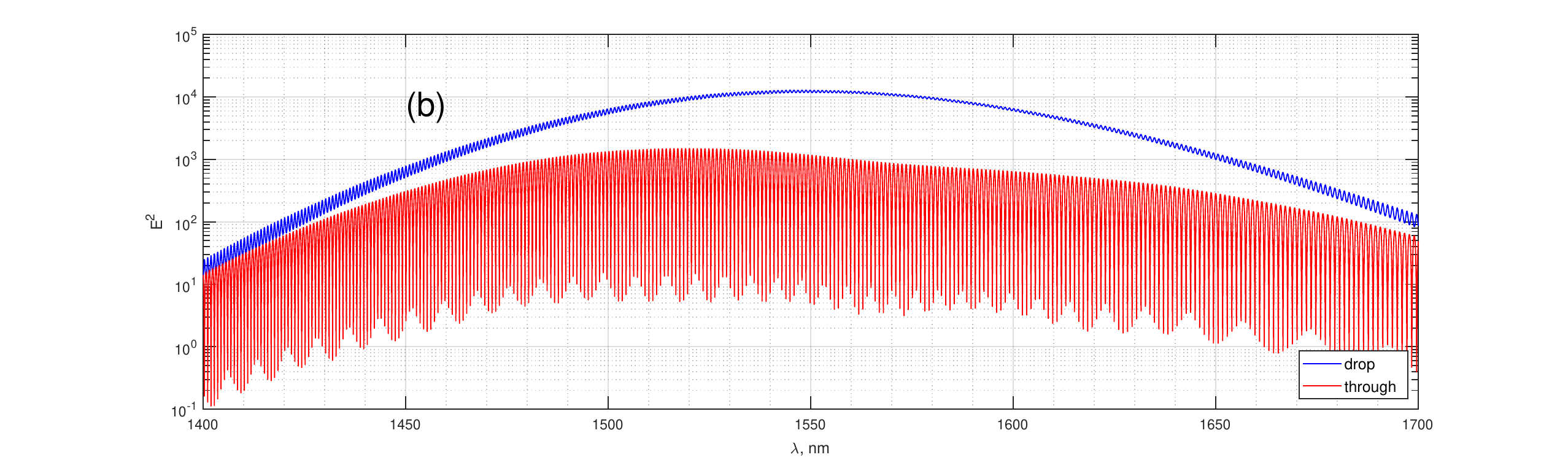}
    \includegraphics[width=0.8\linewidth]{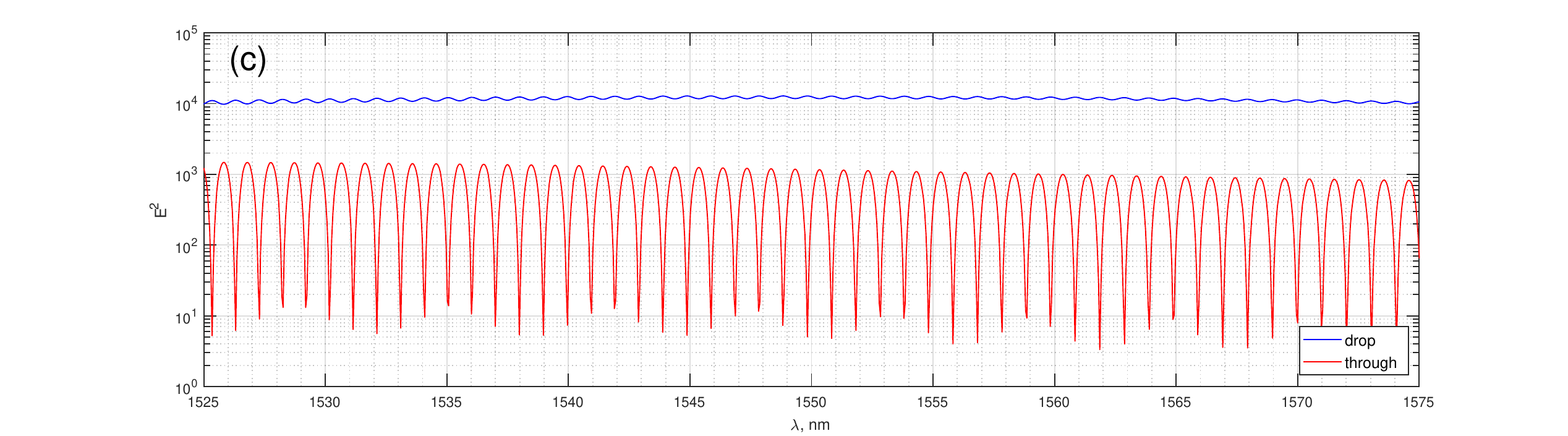}
    \includegraphics[width=0.8\linewidth]{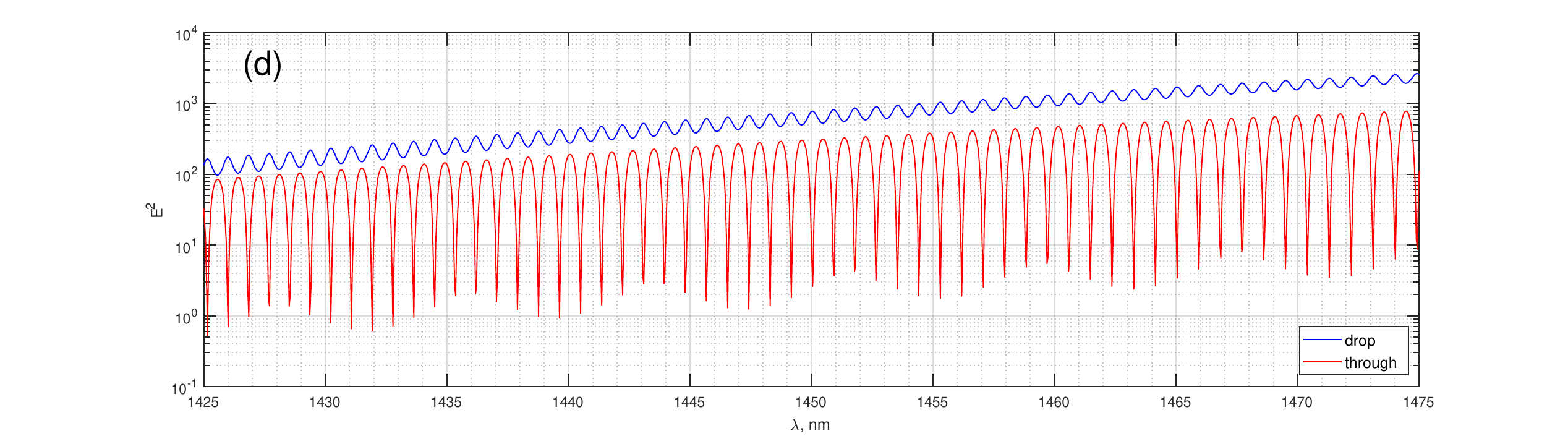}
    
    \caption{Spectra for the \( \sim \) 200 \si{\micro m} radius racetrack resonator/modulator with 8 \si{\micro m} coupling length simulated in ANSYS\textsuperscript{\textregistered} Lumerical varFDTD.}
    \label{fig:12}
\end{figure}

Numerical analysis using ANSYS\textsuperscript{\textregistered} Lumerical varFDTD is used. Additionally, the output mode of an OZ-optics lensed fiber is transferred to a taper optimized for mode coupling. Through and drop port resonances for the \( \sim \)65 \si{\micro m} racetrack resonator/modulator are plotted in Fig. \ref{fig:11}, whereas the through and drop port resonances for the \( \sim \)200 \si{\micro m} radius racetrack resonator/modulator are plotted in Fig. \ref{fig:12}. The spectra from each resonator/modulator exhibit different FSR, such that the relative signals may be indirectly distinguished from their transfer function.

\section{Silicon nitride (SiNx) waveguide edge couplers}
\begin{figure}[H]
    \centering
    \includegraphics[width=\linewidth]{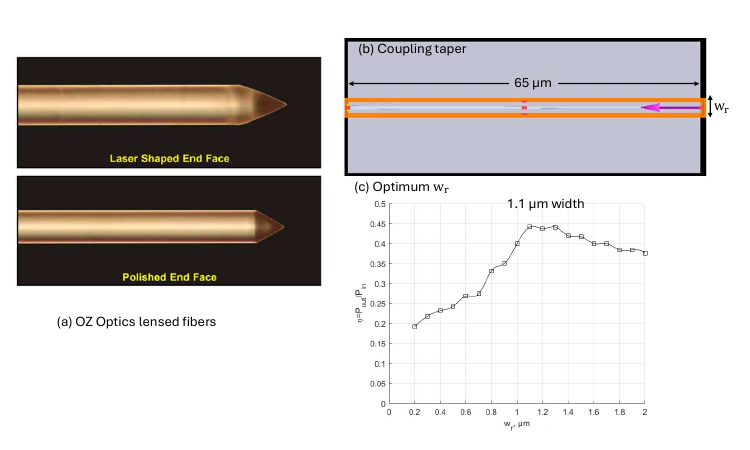}
    \caption{(a) OZ-optics tapered and lensed fibers with a 2.5 \si{\micro m} spot size from a working distance of 14.2 \si{\micro m}. (b) Representation of coupling efficiency in simulation. (c) Optimum coupling width, $\textrm{w}_\textrm{r}$, is \(\sim\)1.1 \si{\micro m} for a 45\% coupling efficiency, $\eta$. Power monitors are used on the input of the taper and on the input of the waveguide, and the efficiency of the coupling calculated via the transmission of power into the waveguide.}
    \label{fig:13}
\end{figure}

\begin{figure}[H]
    \centering
    \includegraphics[width=\linewidth]{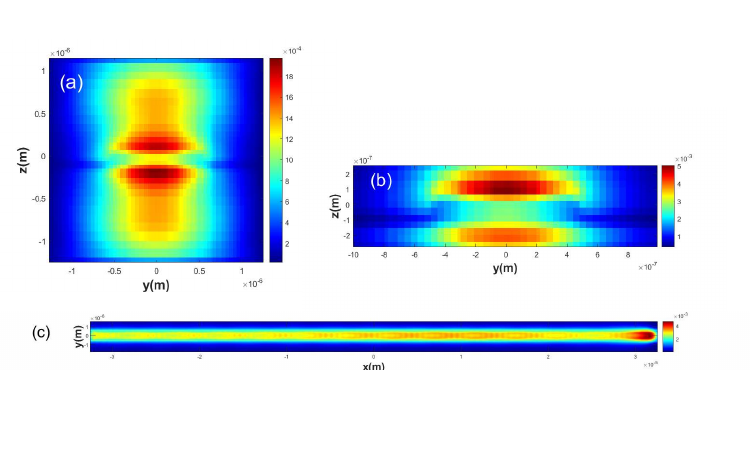}
    \vspace{-8mm}
    \caption{The power in the taper waveguide at the; (a) input port; (b) output port; and (c) central cross-section from the top rectangular prism. }
    \label{fig:14}
\end{figure}

In the case of SiNx, the taper exhibits negligible improvement of mode confinement from the lensed optical fiber. Therefore, the regular waveguide is used with only a slight increase in its width by 100 \si{nm} over 65 \si{\micro m} length, which was optimized via a parametric sweep. Coupling efficiency is calculated based on the power at the end of the taper (start of the waveguide), divided by the total power input to the taper. Figure \ref{fig:13} shows a plot of the SiNx tapered waveguide end as it couples a 2.5 \si{\micro m} spot size beam from an OZ-optics lensed fiber at 14.2 \si{\micro m} working distance. With the use of the taper, the maximum coupling efficiency from the fiber is about 45\%. Fig. \ref{fig:14} shows the relatively coarse optimal power profile at the input of the taper and at the output of the taper (input of the waveguide).

\vspace{-10mm}

\section{Critical dimensions}
The first- and third-stage rings in the third-order ring resonator/modulator share a 200 \si{nm} coupling gap width to their nearest adjacent waveguides. Each silicon ring resonator/modulator in waveguide in the $4\times4$ matrix multiplier shares the narrowest gap of 150 \si{nm}. Otherwise, the gap between rings in the third-order ring resonator/modulator is 358 \si{nm} and the gap between the silicon nitride waveguide and  \( \sim\)62 \si{\micro m} ring is 300 \si{nm} and the \( \sim\)200 \si{\micro m} radius ring 320 \si{nm}. Each of these dimensions are critical to the device performance. In particular, the third-order resonator is optimized with analytical models that may not account for variation in gap width, effective index, or mode confinement. 
At the North edge and South edge of the floor plan, on the Eastern side of the layout, are custom-made edge couplers from SiNx which are terminated 1 \si{\micro m} away from each edge. It is critical that the deep-trench does not scour the tip of each edge coupler, which would prevent signal transmission or introduce significant losses and alignment issues. At the edge of the floorplan, the silicon nitride trench window must extend 20 \si{\micro m} away from the waveguide edge. Therefore, the silicon nitride trench window extends beyond the floorplan. However, in this region the deeptrench may cut the silicon nitride window. It is essential that the deep trench does not incidentally damage or crack the taper. The terminated input/output end of the taper is \( \sim \)1.1 \si{\micro m}, whereas it angles towards the same width of the waveguide \( \sim \)1.0\si{\micro m}. It's width dimension is not critical for efficient coupling to the lensed fiber.

\section{Optical Inputs and Outputs (I/O)}
The optical input/output (I/O) is a transverse-electric (TE) polarized continuous wave of 1550 \si{nm} wavelength, and the inputs are bi-directional six North and South edge couplers as black-box cells. There are a remaining two custom bi-directional edge coupler/tapers that are also SiNx terminators.
Each SiNx taper as a rib waveguide leads into a rib SiNx waveguide. These  edge couplers were simulated to function in ANSYS\textsuperscript{\textregistered} Lumerical FDTD, but have not been tested or measured before. They have a critical dimension with the edge of the floor plan (1 \si{\micro m} from the edge). Each is terminated at 1.1 \si{\micro m}. 

ANSYS\textsuperscript{\textregistered} Lumerical varFDTD was used to simulate propagation of a 1500-1600 \si{nm} pulse through the terminator. It maintains signal intensity and mode confinement from the taper at the waveguide.

\section{Conclusion}
In this work, a theoretical and practical output design for a photonic matrix multiplier and three-ring Butterworth filter were implemented in KLayout, and simulated in ANSYS\textsuperscript{\textregistered} Lumerical 2025. It was discovered through numerical analysis that idealized analytic models may not account for imperfections in the AMF foundry process, or specific design constraints of the Advanced Micro Foundry (AMF) process. Nonetheless, the design achieves acceptable full spectral range (FSR) in varFDTD and 3D FDTD simulation, therefore achieving the desired performance characteristic. 

\section{Recommendation and Disclaimer}
Whereas fabrication tolerances are likely to contribute to a shift in the resonance spectrum of the ring resonator array, post-fabrication annealing through multi-shot infrared femtosecond tuning of the waveguide structures are not recommended due to induced round-trip losses. Alternative methods include cladding index modification, and ion implantation.

\section*{Acknowledgments}
Work on this design layout was was supported by the Canadian Microsystems Passive Silicon Photonics Fabrication Workshop (CMC SiEPIC), and required use of a non-disclosed Advanced Micro Foundry and University of British Columbia Process Design Kit (PDK). The photonic integrated circuit layouts were designed using KLayout and fabricated via CMC Microsystems using the AMF silicon photonics platform. In accordance with foundry non-disclosure policies, specific proprietary process design kit (PDK) layer numbers, internal routing rules, and proprietary sub-cell designations have been abstracted to functional descriptions. The author thanks Vien Van for helpful preliminary discussions regarding initial resonator parameter estimates. Special thanks to Robin Kim, Ruoheng Zhang, Jessica Zhang, and Lukas Christowski for valuable instruction and feedback on practical design constraints. 
\newpage
\bibliographystyle{unsrt}  
\bibliography{references}

@article{canciamilla2012photo,
  title={Photo-induced trimming of chalcogenide-assisted silicon waveguides},
  author={Canciamilla, Antonio and Morichetti, Francesco and Grillanda, Stefano and Velha, Philippe and Sorel, Marc and Singh, Vivek and Agarwal, Anu and Kimerling, Lionel C and Melloni, Andrea},
  journal={Optics express},
  volume={20},
  number={14},
  pages={15807--15817},
  year={2012},
  publisher={Optical Society of America}
}

@article{de2020laser,
  title={Laser trimming of the operating wavelength of silicon nitride racetrack resonators},
  author={De Paoli, Greta and Jantzen, Senta L and Bucio, Thalia Dominguez and Skandalos, Ilias and Holmes, Christopher and Smith, Peter GR and Milosevic, Milan M and Gardes, Frederic Y},
  journal={Photonics Research},
  volume={8},
  number={5},
  pages={677--683},
  year={2020},
  publisher={Chinese Laser Press and Optical Society of America}
}

@article{kalashnikov2022tuning,
  title={Tuning of silicon nitride micro-cavities by controlled nanolayer deposition},
  author={Kalashnikov, Dmitry A and Alagappan, Gandhi and Hu, Ting and Lim, Nelson and Leong, Victor and Png, Ching Eng and Krivitsky, Leonid A},
  journal={Scientific Reports},
  volume={12},
  number={1},
  pages={15074},
  year={2022},
  publisher={Nature Publishing Group UK, London}
}

@article{wu2005femtosecond,
  title={Femtosecond laser absorption in fused silica: Numerical and experimental investigation},
  author={Wu, Alexander Q and Chowdhury, Ihtesham H and Xu, Xianfan},
  journal={Physical Review B—Condensed Matter and Materials Physics},
  volume={72},
  number={8},
  pages={085128},
  year={2005},
  publisher={APS}
}

@article{zoubir2006laser,
  title={Laser-induced defects in fused silica by femtosecond IR irradiation},
  author={Zoubir, Arnaud and Rivero, Clara and Grodsky, Rachel and Richardson, Kathleen and Richardson, Martin and Cardinal, Thierry and Couzi, Michel},
  journal={Physical Review B—Condensed Matter and Materials Physics},
  volume={73},
  number={22},
  pages={224117},
  year={2006},
  publisher={APS}
}

@article{oosterbeek2016effects,
  title={Effects of dopant type and concentration on the femtosecond laser ablation threshold and incubation behaviour of silicon},
  author={Oosterbeek, Reece N and Corazza, Carsten and Ashforth, Simon and Simpson, M Cather},
  journal={Applied Physics A},
  volume={122},
  number={4},
  pages={449},
  year={2016},
  publisher={Springer}
}

@article{zhang2023study,
  title={A study of incubation effects in femtosecond laser irradiation of silicon and copper: R. Zhang et al.},
  author={Zhang, Ruoheng and McDowell, Aran and Hegmann, Frank and Fedosejevs, Robert and Tsui, Ying Yin},
  journal={Applied Physics A},
  volume={129},
  number={2},
  pages={131},
  year={2023},
  publisher={Springer}
}

@article{warner2025simulation,
  title={Simulation of silicon ridge waveguide enhanced two-photon absorption from femtosecond pulses},
  author={Warner, Cael and Zhang, Ruoheng},
  journal={arXiv preprint arXiv:2506.18098},
  year={2025}
}

@phdthesis{zhang2024precision,
  author       = {Zhang, Ruoheng},
  title        = {Precision tuning of silicon photonic devices using multi-shot femtosecond laser irradiation and characterization of incubation effect in multi-shot femtosecond laser irradiated silicon},
  school       = {University of Alberta},
  year         = {2024},
  address      = {Edmonton, AB, Canada},
  url          = {https://ualberta.scholaris.ca/items/bcb2fc13-f1b7-4ace-b814-e2419ab40bb4}
}

@article{zhang2024multi,
  title={Multi-shot near-infrared femtosecond laser tuning of silicon microring resonators},
  author={Zhang, Ruoheng and Leng, Bo and Song, Hanfa and Hegmann, Frank and Fedosejevs, Robert and Van, Vien and Tsui, Ying Yin},
  journal={Optics Communications},
  volume={560},
  pages={130446},
  year={2024},
  publisher={Elsevier}
}

@unpublished{zhanghigh,
  author       = {Zhang, Ruoheng and Leng, Bo and Fedosejevs, Robert and Van, Vien and Tsui, Ying Y.},
  title        = {High-precision tuning of silicon photonic devices by the incubation effect in multi-shot femtosecond laser irradiation},
  note         = {Optica Open Preprint},
  year         = {2026},
  month        = {June},
  doi          = {10.1364/opticaopen.32825879},
  url          = {https://preprints.opticaopen.org/articles/preprint/High-precision_tuning_of_silicon_photonic_devices_by_the_incubation_effect_in_multi-shot_femtosecond_laser_irradiation/32825879}
}

@article{bogaerts2020programmable,
  title={Programmable photonic circuits},
  author={Bogaerts, Wim and P{\'e}rez, Daniel and Capmany, Jos{\'e} and Miller, David AB and Poon, Joyce and Englund, Dirk and Morichetti, Francesco and Melloni, Andrea},
  journal={Nature},
  volume={586},
  number={7828},
  pages={207--216},
  year={2020},
  publisher={Nature Publishing Group UK, London}
}

@book{madsen1999optical,
  title={Optical filter design and analysis},
  author={Madsen, Christi K and Zhao, Jian H},
  volume={3},
  year={1999},
  publisher={Wiley Online Library}
}

@article{bachman2016postfabrication,
  title={Postfabrication phase error correction of silicon photonic circuits by single femtosecond laser pulses},
  author={Bachman, Daniel and Chen, Zhijiang and Wang, Christopher and Fedosejevs, Robert and Tsui, Ying Y and Van, Vien},
  journal={Journal of Lightwave Technology},
  volume={35},
  number={4},
  pages={588--595},
  year={2016},
  publisher={IEEE}
}

@article{callahan_simple_2020,
  title = {A simple model to scope out parameter space for indirect drive designs on {NIF}},
  volume = {27},
  issn = {1070-664X},
  url = {https://doi.org/10.1063/5.0006217},
  doi = {10.1063/5.0006217},
  number = {7},
  urldate = {2025-06-11},
  journal = {Physics of Plasmas},
  author = {Callahan, D. A. and Hurricane, O. A. and Kritcher, A. L. and Casey, D. T. and Hinkel, D. E. and Opachich, Y. P. and Robey, H. F. and Rosen, M. D. and Ross, J. S. and Rubery, M. S. and Young, C. V. and Zylstra, A. B.},
  month = jul,
  year = {2020},
  pages = {072704},
}

@techreport{UAlberta2025_003,
  author      = {{University of Alberta Technology Transfer Services}},
  title       = {Method for Post-Fabrication Trimming of Silicon Photonic Integrated Circuits Using Multi-Shot Laser Irradiation},
  institution = {University of Alberta},
  number      = {ID\# 2025-003},
  year        = {2025},
  type        = {Technology Transfer Summary}
}

@article{beranek2023melting,
  title={On the melting thresholds of semiconductors under nanosecond pulse laser irradiation},
  author={Ber{\'a}nek, Ji{\v{r}}{\'\i} and Bulgakov, Alexander V and Bulgakova, Nadezhda M},
  journal={Applied Sciences},
  volume={13},
  number={6},
  pages={3818},
  year={2023},
  publisher={MDPI}
}

@article{schriever2012strained,
  title={Strained silicon photonics},
  author={Schriever, Clemens and Bohley, Christian and Schilling, J{\"o}rg and Wehrspohn, Ralf B},
  journal={Materials},
  volume={5},
  number={5},
  pages={889--908},
  year={2012},
  publisher={MDPI}
}

@article{yu2022ge,
  title={Ge ion implanted photonic devices and annealing for emerging applications},
  author={Yu, Xingshi and Chen, Xia and Milosevic, Milan M and Shen, Weihong and Topley, Rob and Chen, Bigeng and Yan, Xingzhao and Cao, Wei and Thomson, David J and Saito, Shinichi and others},
  journal={Micromachines},
  volume={13},
  number={2},
  pages={291},
  year={2022},
  publisher={MDPI}
}

\end{document}